\documentclass{aastex}
\usepackage{emulateapj5}

\def\chandra{{\it Chandra}}

\def\asca{{\it ASCA}}
\def\hst{{\it HST}}

\def\merlin{{\it MERLIN}}
\def\vla{{\it VLA}}

\def\lum{erg s$^{-1}$}

\def\nh{cm$^{-2}$}
\def\arcsec{$^{\prime\prime}$}
\def\deg{$^{\circ}$}

\def\ltsima{$\; \buildrel < \over \sim \;$}
\def\simlt{\lower.5ex\hbox{\ltsima}} 
\def\gtsima{$\; \buildrel > \over \sim \;$}
\def\simgt{\lower.5ex\hbox{\gtsima}} 

\begin{document}
\title{Deep Chandra and multicolor HST follow-up of the jets in two powerful 
radio quasars}


\author{Rita M. Sambruna\altaffilmark{1}, Mario Gliozzi, Davide Donato}

\altaffiltext{1}{Current permanent address: Goddard Space
Flight Center, Code 661, Greenbelt, MD 20771}
\affil{George Mason University, Dept. of Physics and Astronomy and School of
Computational Sciences, MS 3F3, 4400 University Drive, Fairfax, VA 22030
(rms@milkyway.gsfc.nasa.gov)}

\author{L. Maraschi and F. Tavecchio}
\affil{Osservatorio Astronomico di Brera, via Brera 28, 20121 Milano, Italy}

\author{C.C. Cheung\altaffilmark{2}}

\altaffiltext{2}{Jansky Postdoctoral Fellow; National Radio Astronomy
Observatory. Now hosted by Kavli Institute for Particle Astrophysics and
Cosmology, Stanford University, Stanford, CA 94305}
\affil{MIT Kavli Institute for Astrophysics \& Space
Research, 77 Massachusetts Ave., Cambridge, MA 02139, USA}

\author{C. Megan Urry}
\affil{Yale University, New Haven, CT 06520}

\author{J.F.C. Wardle} 
\affil{MS 057, Department of Physics, Brandeis University, Waltham, MA
02454}

\begin{abstract} 

We present deep (70--80~ks) \chandra\ and multicolor \hst\ ACS images
of two jets hosted by the powerful quasars 1136--135 and 1150+497,
together with new radio observations. The sources have an FRII
morphology and were selected from our previous X-ray and optical jet
survey for detailed follow up aimed at obtaining better constraints on
the jet multiwavelength morphology and X-ray and optical spectra of
individual knots, and to test emission models deriving physical
parameters more accurately. All the X-ray and optical knots detected
in our previous short exposures are confirmed, together with a few new
faint features. The overlayed maps and the emissivity profiles along
the jet show good correspondence between emission regions at the
various wavelengths; a few show offsets between the knots peaks of
$<$1\arcsec. In 1150+497 the X-ray, optical, and radio profiles
decrease in similar ways with distance from the core up to $\sim$
7\arcsec, after which the radio emission increases more than the X-ray
one. No X-ray spectral variations are observed in 1150+497. In
1136--135 an interesting behavior is observed, whereby, downstream of
the most prominent knot at $\sim$ 6.5\arcsec\ from the core, the X-ray
emission fades while the radio emission brightens. The X-ray spectrum
also varies, with the X-ray photon index flattening from $\Gamma_x
\sim 2$ in the inner part to $\Gamma_x \sim 1.7$ to the end of the
jet. We intepret the jet behavior in 1136--135 in a scenario where the
relativistic flow suffers systematic deceleration along the jet, and
briefly discuss the major consequences of this scenario. The latter is
discussed in more detail in our companion paper (Tavecchio et
al. 2005a).

\end{abstract} 


\section{Introduction} 

There is general consensus that multiwavelength imaging of extended
radio jets in radio-loud Active Galactic Nuclei (AGN) is key to
understanding their physical properties, such as radiative processes,
kinematics, and flow dynamics. The advent of the \chandra\ X-ray
Observatory was a crucial step forward, as for the first time detailed
imaging spectroscopy studies became possible at X-ray wavelengths.
Indeed, \chandra\ has changed our perception of jets.  Once thought to
be rare, X-ray jets are now recognized to be a common phenomenon. Over
its five years of operations, \chandra\ detected bright X-ray emission
from several large-scale (kpc to Mpc) radio jets in both FRI and FRII
radio galaxies (see the on-line catalog at http://hea-www.harvard.edu/XJET/).

An excellent laboratory to study jet physics is provided by the
powerful jets in FRII sources. During \chandra\ cycle 2, we performed
a ``snapshot'' survey at X-ray and optical wavelengths of a sample of
16 FRII jets and 1 FRI jet selected from radio images, without
knowledge of their optical and X-ray properties (Sambruna et al. 2002,
2004; hereafter S02, S04, respectively). The survey demonstrated that
high-energy emission from radio jets is common, with a rate of 60\% of
detections at both X-ray and optical wavelengths (independently
confirmed by Marshall et al. 2005). The multiwavelength morphologies
and the SEDs were consistent with a scenario where the
radio-to-optical emission is due to synchrotron radiation, while the
X-ray emission is mostly due to Inverse Compton (IC) scattering of the
Cosmic Microwave Background photons by very low-energy electrons
(IC/CMB).  In the comoving jet frame, the dominant source of target
photons is provided by the CMB, whose radiation density U$_{CMB}
\propto \Gamma^2(1+z)^4$, with $\Gamma$ the bulk Lorentz
factor of the jet plasma. A major implication of this model is that
the jet is still relativistic on kpc and larger scales (Tavecchio et
al. 2000; Celotti et al. 2001). 

A clean diagnostic of the IC/CMB process and better constraints on the
physical parameters are offered by the X-ray and optical spectra. In
general, X-ray spectra with a photon index similar or flatter than the
radio one are expected in the IC case, while steeper slopes would be a
signature of synchrotron emission. In the IC/CMB scenario, the optical
spectrum constrains both the minimum and the maximum energies of the
particle distribution: $\gamma_{min}$, which dominates total energy
density, and $\gamma_{max}$, which depends on acceleration and
determines the particle lifetime. Unfortunately, the short ACIS-S
exposures and the single-filter, limited sensitivities of STIS of the
cycle 2 observations were unable to provide accurate spectra for
various knots.

In addition to spectra, more detailed X-ray and optical morphologies
are necessary for a better understanding of jet physics. Indeed, an
interesting result of the GO2 survey was the finding that the
X-ray-to-radio flux ratio generally decreases along the jet (S04), as
previously observed in 3C~273 (Sambruna et al. 2001; Marshall et
al. 2001). We interpreted this result in terms of plasma deceleration
at large distances from the core with an increase in magnetic field
due to plasma compression (S04). This scenario was independently
developed by Georganopoulos \& Kazanas (2004). Detailed jet
morphologies at the various wavelengths can provide further insights
into the plasma deceleration scenario.

Thus, we selected two jets for deeper \chandra\ and more sensitive
multi-color \hst\ ACS follow-up observations, 1136--135 ($z$=0.558)
and 1150+497 ($z$=0.334). The observations were designed to achieve
sufficient signal-to-noise to determine the X-ray and optical
continuum slopes and jet morphologies accurately within reasonable
exposures. In this paper and in a companion one (Tavecchio et
al. 2005a; Paper II in the following) we concentrate on the
multiwavelength jet properties and their interpretation. Results from
the analysis of the X-ray cores are discussed in a separate
publication.

The structure of the paper is as follows. In \S~2 we describe the
targets and in \S~3 the observations. The results are discussed in
\S~4 and the interpretation in \S~5. Discussion and summary follow in
\SS~6 and 7.  

Throughout this work, a concordance cosmology\footnote[1]{In S02, S04
we used a standard cosmology with H$_0=75$ km s$^{-1}$ Mpc$^{-1}$ and
$q_0=0.5$. To compare the luminosities of the present work with the
previous papers, divide the luminosities by 1.695 for 1135--135 and
1.489 for 1150+497.} with H$_0=71$ km s$^{-1}$ Mpc$^{-1}$,
$\Omega_{\Lambda}$=0.73, and $\Omega_m$=0.27 (Bennett et
al. 2003). With this choice, 1\arcsec\ corresponds to 6.4~kpc for
1136--135 and 4.8~kpc for 1150+497. The energy index $\alpha$ is
defined such that $F_{nu} \propto \nu^{-\alpha}$.

\section{Targets and Previous Jet Observations} 

The targets, 1136--135 and 1150+497, were selected from our \chandra\ and
\hst\ GO2 survey (S02, S04) because of their representative
properties. In the case of 1136--135, the X-ray jet displayed an
interesting mix of synchrotron and IC/CMB emission, with the former
process suspected to dominate in the inner knot (knot A) detected in the
10 ks X-ray image, and the IC/CMB dominating in the outer knots.  The
case of 1150+497 is our brightest example of a jet whose X-ray emission
could derive entirely from the IC/CMB process, with well-resolved X-ray
knots and an interesting twisted morphology. 

The basic properties of the targets are summarized in Table~1. The
radio and optical data were taken from the literature (Liu \& Zhang
2002 and \verb+NED+, respectively). Below we summarize the results
from our GO2 snapshot \chandra\ and \hst\ observations (S02, S04).

\noindent{\bf 1136--135}: The short-exposure 
ACIS-S map of the jet shows two X-ray knots at 4.5\arcsec\ and
6.7\arcsec\ from the nucleus, both with a bright optical counterpart
in our \hst\ images.  Based on the shape of the SED, the X-ray
emission from the innermost knot was interpreted as due to synchrotron
from the high energy tail of the electron population emitting the
radio continuum, while IC/CMB was the favored mechanism for the X-rays
from the 6.7\arcsec\ knot; these mechanisms account also for the
X-ray/radio morphology of this jet. 
From the three-point SED, we derived Doppler factor $\delta \sim 7$,
magnetic field $B \sim 10-40~\mu$Gauss, and total kinetic jet power
P$_{jet} \sim 5 \times 10^{46}$ \lum.

\noindent{\bf 1150+497}: This 10\arcsec-long jet is the brightest X-ray
jet in our survey and a case of ``pure'' IC/CMB emission. At X-rays there
are at least three bright knots at 2.1\arcsec, 4.3\arcsec, and
7.9\arcsec\ from the core.  The latter coincides with a terminal hot spot
in the radio image. All three knots were detected in the \hst\ image.
Interestingly, this X-ray jet has a ``twisted'' morphology with a total
change in position angle of $\Delta PA \sim 20$ degrees, following
closely the radio (Owen \& Puschell 1984). This suggests that the beaming
may change along the jet as a result of the wiggling of the plasma. In
this scenario, because of the different Doppler boosting chracterizing
the different regions of the jet, the most intense emission comes from
those portions of the jet whose velocity is close to the line of
sight. The SEDs of the two inner knots (S02) are both consistent with
IC/CMB emission at X-rays, with $\delta \sim 6$, $B \sim 25~\mu$Gauss,
and P$_{jet} \sim 3 \times 10^{47}$ \lum.

The follow-up observations, described below, were designed in order to
acquire accurate X-ray and optical spectra of the individual bright
knots, to test emission models for the SEDs.

\section{Observations}  

\subsection{Chandra}

The \chandra\ observations were carried out on 2003 April 16 for
1136--135 and on 2003 July 18 for 1150+497 (obsid 3973 and 3974), with
total exposures of 77.4~ks and 68~ks, respectively. Both datasets were
obtained with ACIS-S with the sources at the aim-point of the S3 chip.
Since we expected bright X-ray cores, we used $\frac{1}{8}$ subarray
mode to reduce the effect of pileup of the nucleus, with an effective
frame time of 0.44~s. With this precaution, the core of 1136--135 had
0.61 cts/frame, corresponding to a small pileup (4\%); the core of
1150+497 had 1.25 cts/frame, or 11\% pileup.  In addition, for each
source, a range of roll angles was specified in order to locate the
jet away from the charge transfer trail of the nucleus and avoid flux
contamination.

The \chandra\ data were reduced following standard screening criteria
and using the latest calibration files provided by the \chandra\ X-ray
Center. The latest version of the reduction software \verb+CIAO+
v. 3.1 was used. Pixel randomization was removed for imaging purposes
only, and only events for \asca\ grades 0, 2--4, and 6 and in the
energy range 0.3--8 keV 
were retained.  We also checked that no flaring background events
occurred during the observations.  After screening, the effective
exposure times are 70.2~ks for 1136--135 and 61.7~ks for 1150+497. The
X-ray images of the jets are shown in Figures~1 and 2, top-left panels
while the X-ray count rates for the various knots are listed in Table~2. 

Background spectra and light curves were extracted from source-free
regions on the same chip of the source. For the jets knots the spectra
were extracted from ellipses of axes in the range 0.5--1\arcsec,
depending on knot's size and location, centered on the position of the
knot's radio counterpart. An important issue, further discussed in
\S~4.3, is the presence of a non-zero basline in the X-ray emissivity
profile of the jets (Figures 5 and 6). While some knots stand out
clearly against this baseline (e.g., A in 1136--135 and E in
1150+497), most are fainter and blend with the surrounding regions. We
thus used extraction regions of different dimensions for the various
knots, depending on their FWHMs. In all cases, the size of the
extraction region in both dimensions (parallel and perpendicular to
the jet's axis) is \gtsima the measured FWHM of the knot. The fluxes
(but not the count rates) were corrected for the effects of a finite
``aperture''. 

The response matrices were constructed using the corresponding thread
in \verb+CIAO+ 3.1. The ACIS spectra with $>$ 100 counts were analyzed
in the energy range 0.5--8 keV, where the calibration is best known
(Marshall et al. 2004) and the background negligible, and were grouped
so that each new bin had $\ge$ 20 counts to enable the use of the
$\chi^{2}$ statistics. The spectra were fitted within \verb+XSPEC+
v.11.2.0. Errors quoted throughout are 90\% for one parameter of
interest ($\Delta\chi^2$=2.7). The unbinned X-ray spectra with $<$ 100
counts were fitted using the C-statistics.

The 0.3--8 keV counts of the sources, extracted from the regions
described above, are reported in Table~2 for each knot. Note that the
counts were extracted from different-size regions for each knot. This
was necessary to avoid overlaps in the extraction regions for
contiguous knots at the limited \chandra\ resolution. Following the
radio, elliptical extraction regions were used in the X-rays.

\subsection{HST} 

Both 1136--135 and 1150+497 showed optical counterparts to their
radio/X-ray jets in our earlier STIS images (S02, S04).  In order to
constrain the optical spectrum of the brightest optical features, we
observed the two objects in three filters with the Advanced Camera for
Surveys (ACS) aboard \hst.   
The total exposures were one orbit for 1136--135, which has the
brighter optical knots, and two orbits for 1150+497 which is
relatively fainter. The central frequencies of the three filters,
F475W (SDSS g), F625W (SDSS r), and F814W (broad I), are 6.32, 4.75,
and 3.72 $\times 10^{14}$ Hz, respectively. Longer total integrations
were obtained at the shorter wavelengths because of the expected steep
spectrum nature of the jet emission. In the case of the F814W exposure
of 1136--135, we received two consecutive segments of different duration
each (338s and 182s). 

The data were obtained from the STScI archive with pipeline processing
applied. The pipeline implements the MultiDrizzle (Koekemoer et
al. 2002) program to deliver calibrated and cosmic-ray cleaned images.
Accurate photometry of the faint jet knots is limited by the high
background levels due to sensitivity of the ACS to field sources, and
contamination from the central sources, in the form of scattered light
and prominent diffraction spikes from the bright quasar nucleus and
diffuse emission possibly from the host galaxy. To isolate the jet
knot emissions, we used square apertures centered on the
X-ray/radio aperture positions. The elliptical apertures we defined
for the \chandra/radio data are not practical to use due to the
aforementioned contamination sources.

In the case of F814W images of 1136--135, the MultiDrizzle pipeline could
not remove cosmic rays. Thus, we adopted the following procedure: we
used the longer segment to remove the host galaxy's contribution and
to measure the jet fluxes. For knots C and D, where cosmic rays
contaminate the extraction region for the fluxes, we used the shorter
segment which was clear.

We estimated the background using identical apertures (9x9 image pixels)
in adjacent regions to the extractions.  
Several background regions at different positions were used; since the
background varies, we used the average and standard deviation of the
values estimated in the boxes. The net flux was derived as the
difference of the total flux minus the average background. As
uncertainty on the net source flux, we used the error percentage
defined as the ratio of the background standard deviation over the net
source flux.

We claim detections in the cases where the net source flux is at least
3$\sigma$. For the non-detections, we quote 3$\sigma$ upper limits. 
Counts were converted to flux densities using the inverse sensitivity
measurement given by the keyword PHOTFLAM in the image headers.   
Extinction corrections were estimated using values for the closest
matching standard bands provided in NED (from Schlegel et
al. 1998). These corrections ranged from a few up to 17$\%$ at the
shortest wavelength. The fluxes in each filter are listed in Table~3
for each knot. 

\subsection{Radio} 

As part of our \chandra\ and \hst\ survey (S02, S04) we obtained
multi-band radio data for the jets which had X-ray and/or optical
detections.  Here, we use the new data obtained with the
NRAO\footnote[2]{The National Radio Astronomy Observatory is a
facility of the National Science Foundation, operated under a
cooperative agreement by Associated Universities Inc.} Very Large
Array at 22 GHz, and with the \merlin\footnote[3]{\merlin\ is a
National Facility operated by the University of Manchester at Jodrell
Bank Observatory on behalf of PPARC} array at 1.7 GHz both achieving
$\sim$0.15--0.3\arcsec\ resolution. The 5 GHz \vla\ data were
previously presented in our GO2 papers (S02, S04).

The new data are discussed in full elsewhere (Cheung 2004; Cheung et
al. in preparation).  Briefly, 22 GHz \vla\ data were obtained in
2002, May and July for 1136--135 and 1150+497, respectively, with
integrations of about three hours on each target (1$\sigma$ rms
$\sim$0.1 mJy/bm). The B-configuration was used for 1150+497 to
achieve similar resolution to the 5 GHz, A-configuration image (Owen
\& Puschell 1984; Fig.~4). The 1136--135 observations utilized the hybrid BnA
configuration to achieve a more circular beam on this low declination
target. The \merlin\ 1.7 GHz observations were obtained in Jan-Feb
2002 over several observing runs (rms $\sim$ 0.3--1.0 mJy/bm). Initial
calibration was performed at Jodrell Bank by the archivists and the
final imaging and self-calibration done at Brandeis.

In addition, a new full-track 8.5 hour, 8.4 GHz \vla\ observation (8 Nov
2003) of 1136--135 was obtained for the purpose of this work. This was
necessary because the other radio images did not detect the inner
($<$5\arcsec) radio jet clearly whereas the \chandra\ image did. The
on-target time was 7.7 hrs with the remaining time spent on calibrator
sources. The resultant rms noise in the naturally weighted image is
approximately 10$\mu$Jy, a few times what is expected from thermal
noise, resulting in a very high-dynamic range image (Figures~3-4).

The radio fluxes for the jet regions were extracted using identical
elliptical apertures to those used in the X-ray analysis. Errors in
flux density are, conservatively, 10\% for the brightest features,
with larger values estimated for fainter features and the noisier (1.7
and 22 GHz) data. Fluxes are reported in Table~3. 

The main purpose of our new deep \vla\ 8.46 GHz map of 1136--136 was
to image/detect the faintest features in its jet (knots $\alpha$ and
A). The emission from these knots were not well-detected in the other
data (\merlin\ 1.7 GHz, archival snapshot \vla\ 5 GHz, and new 22 GHz)
and their fluxes would be underestimated from these maps. The knots,
however, appear fairly well in our reprocessed \vla\ 1.5 GHz map from
the Saikia et al. (1990) data and we were able to estimate their
fluxes above the surrounding extended lobe emission. The fluxes are:
10.9 $\pm$ 1.6 mJy for knot $\alpha$, and 12.1 $\pm$ 1.8 mJy for knot
A. Using these fluxes, and those at 8.46 GHz, the radio energy indices
were derived for these two knots. They are consistent with those
measured for the brighter regions of the jet (Table~3). The 
radio fluxes of knots $\alpha$ and A at 5 GHz and
22 GHz reported in Table~3 were extrapolated assuming these
energy indices and the 8.46 GHz flux.  

\section{Results} 

\subsection{Imaging} 

Figures~1-2 show the \chandra\ images of 1136--135 and 1150+497,
respectively. The top-left panel is the ACIS-S raw image in the total
energy range 0.3--8 keV, with no smoothing applied. The remaining
panels show the smoothed ACIS-S images in total (0.3--8~keV), soft
(0.3--2~keV), and hard (2--8~keV) X-rays. The ACIS-S images were
smoothed using the sub-package {\it fadapt} of {\it FTOOLS} with a
circular top hat filter of adaptive size in order to achieve a minimal
number of 10 counts under the filter. The faint horizontal streak
originating from the cores is the readout streak, caused by
out-of-time events from bright point sources (cores).

The multiwavelength images are presented in Figures~3-4, where the
smoothed 0.3--8~keV ACIS and the \hst\ ACS images are illustrated with
the 8.46~GHz (1136--135) and 4.9~GHz (1150+497) radio contours
overlaid. The optical images are from the F625W filter because of its
higher signal-to-noise ratio and the lower contamination from cosmic
rays, which could not be removed automatically. Optical emission is
present from several of the X-ray and radio knots; the chance
probability that the optical features are due to the coincidental
superposition of background stars or galaxies is \ltsima $2\times
10^{-3}$ for 1136--135 and \ltsima $3\times 10^{-3}$ for 1150+497,
while the combined X-ray/optical probabilities are \ltsima $10^{-5}$
and $10^{-6}$, respectively.  Also shown in Figures~3-4 are the 22 GHz
radio images. While confirming all the previously detected features,
the observations reveal important new details on the jets structure,
as discussed below for each source.

To investigate the jet structure at X-rays we plot in Figures~5-6 the
X-ray profiles of the jet. The latter were extracted from ACIS images
after rebinning (factor 0.2, yielding pixel size 0.1\arcsec) and
smoothing with \verb+fadapt+ with a threshold of 10 counts. For the
extraction region we used a box of width 1.5\arcsec\ and length equal
to the jet length, or 9.4\arcsec\ for 1136--135 and 7.6\arcsec\ for
1150+497. The jet flux was collapsed on the central axis of the
box. Shown in Figures~5-6 are the total 0.3--8~keV, soft (S)
0.3--2~keV, and hard (H) 2--8~keV jet profiles. The bottom panel shows
the run of the softness ratios, defined as (S-H)/(S+H), along the
jets. We stress that these profiles, and those in Figures~7-8, were
derived to show the {\it qualitative} run of the fluxes along the
jet. The uncertainties vary between 10\% and 40\% with the largest
errorbars on the hard X-ray and softness ratio data. Overall, in both
cases the jet emission profile consists of a non-zero baseline over
which more or less defined peaks are superposed.  

The jet profiles at X-rays (total band), optical (F625W), and radio
(8.5~GHz for 1136--135, 4.9~GHz for 1150+497) are illustrated in
Figures~7-8. No rebinning was applied to the radio and ACS data. For
the radio profile we used the same extraction procedure as for the
X-rays. However, for the optical profile multiple and narrower boxes
were used, to follow the jet bending and to limit the background
contamination. The horizontal dotted lines mark the background; the
contribution of the core PSF wings dominates the background in the
inner parts (not plotted). For the optical data, as the background
varies depending on position, we plotted an average value. However, a
few features which appear in Figures~7-8 to be above the plotted
background are in reality only upper limits, due to a higher local
background. These are marked by vertical arrows.

The vertical dashed lines in all panels mark the position of the
knots. A word of caution is necessary in this regard. Only a few X-ray
features stand out clearly and all correspond to well defined radio
features. Towards the end of the jets the X-ray emission appears more
diffuse and the positions of the radio features were used to mark the
knots in Figures~7-8. However, in some cases the radio knots do not
have clear counterparts at the shorter wavelengths (e.g., knot G in
1150+497, see below). In the the jet of 1150+497 the knots appear less
prominent than in 1136--135 and individual fainter knots in the jet
profiles cannot be resolved unambiguously.

Also listed in Table~2 are the FWHM of the X-ray knots in the
direction orthogonal to the jet's axis and in the energy range 0.3--2
keV. The latter were extracted in the softer part of the energy range
where the PSF is better described and less energy-dependent (Karovska
et al. 2001). To measure the FWHM of the knots we used the rebinned
and smoothed ACIS-S images described above. The knot profile in the
direction orthogonal to the jet's axis was extracted by collapsing the
counts into a narrow slice of dimensions 1\arcsec$\times$3\arcsec\
(\verb+projection+ in ds9). Then, the FWHM of the gaussian-like
distribution of the counts was estimated simply by measuring the width
at half maximum.  For the PSF, we used the core image and followed the
same procedure to extract the profile of the point-like core. Note
that weak diffuse X-ray emission around the cores is present (Sambruna
et al. 2005, in prep.); however, the extended emission around the core
becomes non-negligible after a radius of 5\arcsec. Thus, the PSF was
measured inside 3\arcsec\ and it is not affected by the diffuse
emission. The comparison between the PSF and the knot's profile was
used to gauge whether the knot was resolved.

In Table 2 the uncertainty on the FHWMs is \gtsima 0.1\arcsec, which
is the size of the pixel in the images. Also listed is the difference
in quadrature between the PSF FWHM (0.6\arcsec) and the knots'. From
Table 2 it can be seen that, assuming a minimum uncertainty of
0.1\arcsec, only knots $\alpha$ and A in 1136--135, and knots B, C,
and E in 1150+497 are only slightly resolved at 1$\sigma$.  Similar
results are obtained for profiles extracted in both soft (0.3--2 keV)
and hard (2--8 keV) energy bands.

To evaluate the FWHM of the knots in the direction along the jet, we
fitted the knots profiles in Figures 5 and 6 with Gaussians. Ideally,
these profiles should be compared to the core's; in practice, the
presence of non-zero faint emission in between the knots hampers this
comparison. In addition, this procedure can only be performed for
the isolated knots -- i.e., knot A for 1136--135 and knot E (at soft
X-rays) for 1150+497. Both knots are unresolved in the direction
parallel to the jet. 

We now comment on the results for the jets of the two sources
separately.

\subsubsection{1136--135} 

Figures~1 and 3 show that all the X-ray features detected in the
short exposure (S02, S04) are confirmed in this longer
observation. However, more detail is hereby revealed at the larger
signal-to-noise ratio of the ACIS data. 

The innermost part of the jet at 2\arcsec\ from the core (knot
$\alpha$ in S04) is clearly visible. As apparent from Figure~1, knot
$\alpha$ is more prominent at soft X-rays where it is composed of a
faint stream of emission emanating from the core and terminating in a
more diffuse knot. There is a possible optical counterpart in the
\hst\ image (Figure~3), which is clearly detected only after an
appropriate galaxy subtraction. At radio wavelengths, knot $\alpha$
is detected at 4.9 and 8.5 GHz. 

Knots A and B discussed in S02 are confirmed and resolved
(Table~2). Here knot B appears to be more complex than a single
feature, with possible substructures, especially at soft X-rays
(Figures~1 and 5). Unfortunately, the two substructures are too small
(radii $<$ 1\arcsec) preventing a meaningful analysis; in Table~2 and
3 we quote the total X-ray flux from knot B. The count rate is 0.0028
c/s, slightly smaller than 0.005 c/s quoted in S04. The discrepancy is
probably due to the smaller extraction radius used here
($\sim$0.7\arcsec) compared to 1\arcsec\ used in S02, S04. 

Three new X-ray features are identified in the deep ACIS image: knots
C, D, and E at 7.7\arcsec, 8.6\arcsec, and 9.6\arcsec, respectively
(Table~2). Knots C and D have optical counterparts (Figure~7 and
Table~3). The final part of the jet, hotspot HS at 10.3\arcsec, was
formerly known as ``knot C'' in S02, S04.

Inspection of the jet profiles in Figures~5 and 7 shows that there
is a generally good correspondence between knots at the various
wavelengths, but with a few notable exceptions. First, concentrating
on the soft and hard X-ray profiles, knot $\alpha$ is more pronounced
at soft X-rays and barely present at harder energies. Similarly, knot
C and D disappear in the hard X-ray band, while a new component
emerges in between them at $\sim$ 8\arcsec. Interestingly, for knot A
there is an offset of $\sim$ 0.4\arcsec\ ($\sim$ 2 kpc projected size)
between the soft and hard X-ray peaks, with the softer energies
further downstream than the hard X-rays. However, the position of the
X-ray knot in the total band aligns well with the optical and radio
peaks (Figure~7).

Another note of interest is the presence of non-zero inter-knot X-ray
emission especially in the inner portion of the jet (see below). While
it is possible that the fainter inter-knot emission observed in some
X-ray jets (Chartas et al., 2000; Schwartz et al. 2000; Marshall et
al. 2001; S04) is actually due to unresolved faint features, very weak
diffuse emission cannot be excluded. This is discussed in \S~4.2. 

Also interesting is the overall structure of the jet at the three
wavelengths - X-rays, optical, and radio. Figure~7 shows a
qualitatively different jet behavior before and after knot B, the most
prominent feature at the shorter wavelengths. In the X-rays, upstream
of knot B the jet is mainly composed of discrete, isolated peaks,
while downstream of knot B the X-ray emission cascades with
``continuous'' emission and a few faint knots superposed. The radio
profile shows a different behavior. At these wavelengths knot B is
barely discernible, while knots $\alpha$ and A are detected, but
faint. The radio profile increases steadily downstream from knot B and
peaks at the end of the jet, opposite to the X-ray profile. The
optical profile is intermediate, consisting of isolated knots both in
the inner and outer parts.

The multiwavelength profiles of the 1136--135 jet are strongly
reminiscent of the 3C~273 morphology (Sambruna et al. 2001). In the
latter jet, in fact, opposite trends between the runs of the X-ray and
radio profiles were observed similar to the case of 1136--135. This led
us (Sambruna 2001, S04) and other authors (Georganopoulos \& Kazanas
2004) to suggest plasma deceleration as a cause of the observed decrease
of the X-ray to radio ratio from the inner to the outer regions of the
jet. We will discuss this scenario more quantitatively in Paper II.

\subsubsection{1150+497} 

This jet has an interesting wiggling morphology. All the detected
features from S02 are confirmed.

The innermost radio knot A can not be detected even at optical
wavelengths where it is covered by the core PSF. The brightest
detected and resolved feature at both X-rays and optical is knot B at
$\sim$ 2\arcsec. It appears complex especially at soft X-rays, where
it exhibits a faint arc-like structure to the South, ending in knots
C+D (Figure~2). Radio knot E, not previously seen at X-rays, is now
weakly detected and appears to be soft. Similarly, knots F and G, and
hotspot H, previosuly detected in the short GO2 exposures (S04), are
confirmed.

Figure~6 displays the X-ray jet profiles in the soft and hard
bands, while the total X-ray, optical, and radio profiles are shown in
Figure~8. Overall, the run of the flux along the jet is similar at
all wavelengths, in  contrast to 1136--135. The two most
prominent knots, B and E, and the terminal part of the jet, IJ, show a
rough 1:1 correspondence between the X-rays, optical, and radio. Knots
C and D are visible mainly at X-rays and optical (D). For knot G, the
vertical dashed line traces the position of the radio knot; no feature
is present at X-ray and optical at the radio position. However, there
are possible counterparts located $\sim$ 0.5\arcsec\ and $\sim$
0.3\arcsec\ upstream of the radio position at X-rays and optical,
respectively. 

Another interesting feature of the multiwavelength profile is the
sudden increase of the radio emission at the end of the jet. A more
modest rise is  present at X-rays. This behavior is qualitatively
similar to  that of 1136--135 though on a smaller angular scale.
If the terminal feature of this jet is interpreted as a hot-spot 
in the lobe, plasma deceleration is again plausible (see discussion in
\S~5.2). 

\subsection{Spectral Analysis} 

\noindent{\bf X-ray spectra:} 
The X-ray spectra were fitted with a single power law plus Galactic
absorption model, which provided generally a satisfactory fit to the
data. The results of the spectral analysis are reported in Table~2,
where the photon indices, $\Gamma_X$, and 90\% uncertainties are
listed. (The photon index is related to the energy index as
$\Gamma_X=\alpha_X+1$.) The aperture-corrected 1~keV flux densities
are reported in Table~3, together with the fluxes from the other
wavelengths. 

In 1150+497, the X-ray slopes are similar for all the knots, $\Gamma_X
\sim 1.5-1.7$. In 1136--135, the fitted photon indices for the brightest X-ray knots
$\alpha$, A, and B are relatively steep, $\Gamma_X \sim 2.0$, although
the uncertainties are large. The photon index for knot B is marginally
(\ltsima 2$\sigma$) consistent with the value previously quoted in
S04, $\Gamma_X=1.5 \pm 0.5$. The discrepancy is due to the larger
extraction radius used in S02, S04, which includes a contribution from
the following knot C, previously unidentified. In fact, the measured
X-ray slope of knot C at 7.7\arcsec\ is $\Gamma_X \sim 1.5$, flatter
than in knot B. For the remaining knots C, D, E, and the hotspot HS
the X-ray slopes are affected by large uncertainties.

Additionally, we fitted the brighter knots' X-ray spectra with a
thermal bremmstrahlung and a composite model consisting of the thermal
bremmstrahlung plus a power law. The fits with the thermal or thermal
+ power law model are significantly worse than the single power law,
and/or yield unphysical values for the temperature/photon index. 
  
The 1~keV flux densities listed in Table~3 were derived using the
photon indices from the spectral fits. For the weak knots with $<$ 100
counts, however, we calculated the fluxes from the counts in Table~2
assuming $\Gamma_X=1.8$ plus Galactic column density for both
sources. These are fully consistent with the fluxes calculated using
the slopes in Table~3, which are affected by large uncertainties. 

We further investigated X-ray spectral variations along the jets by
means of softness ratio profiles, as shown in the bottom panels of
Figures~5-6. The softness ratios are defined as SR=S-H/(S+H), where S
are the counts in the energy band 0.3--2~keV and H the counts in
2--8~keV. As seen in Figures~5-6, X-ray spectral variability is
clearly present along the jets. In both sources, a trend is apparent
of softer X-ray spectra at the knot positions than in the inter-knot 
regions, however, as stated above the uncertainties are large. 
 
The core X-ray spectra exhibit interesting properties, which will be
presented in a future publication. 

\noindent{\bf Optical spectra:} Table~3 lists the dereddened optical
fluxes for the detected knots in the various filters. Upper limits are
reported for the non-detections. Note that some knots have a firm
detection in only one or two filters. 

To derive the optical slopes, we performed a linear fit to the fluxes
in Table~3. To take into account the optical upper limits for some of
the knots, we used the \verb+ASURV+ statistical package. However, no
reliable estimate of the optical slope was obtained, given the small
number of datapoints available to begin with. 

Thus, we only considered knots for which firm detections in at least
two filters are present. The slopes derived from a linear interpolation
of the optical fluxes are listed in Table~3. No uncertainties are
listed for the knots with only two detections, as errors in these
cases are meaningless. While we report the derived slopes also for the
faintest optical knots (marked in Table~3), these values should be
regarded with caution due to the lower signal-to-noise ratio of the
data. 

\noindent{\bf Radio spectra:}
Also listed in Table~3 are the radio energy indices, derived from a
linear interpolation to the available radio data (see \S~3.3). The
average index is $\alpha_{rad}=0.7$ for both jets. Interestingly, the
radio spectral indices are comparable to the X-rays slopes in the
cases of knots $\alpha$, A, B, C, and D for 1136--135, and all knots for
1150+497. This is expected if the same population of very low energy
electrons is responsible for scattering the CMB photons into the X-ray
energy band.

\subsection{Residual X-ray emission around the knots}  

In S04 we noted the presence of faint residual X-ray emission in
between the knots in several jets. Emission from the inner jet was
also detected in PKS 0637--752 in a 100~ks ACIS-S exposure (Chartas et
al. 2000, Schwartz et al. 2000), in 3C~273 (Marshall et al. 2001), and
in the M87 jet (Perlman \& Wilson 2005).

As shown by visual inspection of Figures~1-2, faint X-ray emission
seems to be present between the knots in both the jets of 1136--135
and 1150+497.  This is confirmed more quantitatively by the jet
profiles in Figures~5-6 which show a non-zero baseline in between the
individual emission peaks, above the ACIS background. For easiness of
reference, we will call this residual X-ray emission ``inter-knot''.

We extracted the spectrum of the inter-knot X-ray emission, by
subtracting the knots's contribution from the total jet spectrum. The
extracted spectrum, with a total of 241 and 326 counts for 1136--135
and 1150+497, respectively, represents the {\it total} inter-knot
X-ray emission.  In the assumption that the emission has the same
origin everywhere throughout the jet (an oversimplistic assumption),
the X-ray spectrum obtained with this procedure was fitted using a
thermal bremsstrahlung and a power-law models.

Despite the relatively large number of counts the spectra are 
quite noisy, warranting the use of the C-statistics in the spectral
analysis. The noiseness of the spectra could be related to the large
extraction region, encompassing a wide physical region of the jet
where multiple emission mechanisms and/or parameter gradients could be
present. From the fits, we find that it is not formally possible to
distinguish between a thermal and non-thermal origin of the X-ray
emission from the inter-knot region in 1150+497: both the
bremmstrahlung and a power-law models represent the data adequately in
this jet. In the case of 1136--135 the bremmstrahlung is better
($\Delta$Cstat=10 for same d.o.f.s) than the single power law. The
fitted temperature is very similar in both sources,
$kT=2.7^{+1.4}_{-1.0}$ keV for 1136--135, and $kT=2.7^{+1.7}_{-0.7}$
keV for 1150+497. The power-law photon indices are
$\Gamma_x=1.6^{+0.5}_{-0.1}$ for 1136--135, and $\Gamma_x=1.9 \pm 0.2$
for 1150+497.  The observed 0.3--8~keV fluxes are $1.6\times
10^{-14}{\rm~erg~cm^{-2}~s^{-1}}$ and $2.8\times
10^{-14}{\rm~erg~cm^{-2}~s^{-1}}$ for 1136--135 and 1150+497
respectively, and the corresponding intrinsic luminosities are
$3.1\times 10^{43}{\rm~erg~s^{-1}}$ and $1.3\times
10^{43}{\rm~erg~s^{-1}}$.

The origin of the inter-knot X-ray emission is not clear. A first
possibility includes thermal emission from shocked ambient gas as the
plasma plows its way through the ISM. However, there is no evidence of
such extended emission in the radio images, where the knots appear
elongated in the direction of the jet's axis (Figures 3 and
4). Another argument against thermal emission is the lack of rotation
measure in the radio (\S~5.3). 

An alternative scenario is that the faint inter-knot X-ray emission is
due to unresolved features (e.g., oblique shocks or a population of
hot electrons, see \S~5.3), or simply are due to our definition of
``knot'', forcefully arbitrary and resolution-dependent. Indeed, on
one side in order to quantify their emission we are forced to assume a
finite size for the knots (\gtsima the PSF size); on the other, it is
unlikely that the physical emission region is intrinsically well
defined, for example due to particle diffusion and escape, or spatial
and/or temporal inhomogeneities. A related problem is the faintness of
the emission at the end of the jet, where the emission peaks are less
well separated from each other (Figure 5 and 6).

We conclude that there is evidence for the presence of unresolved
X-ray emission around the knots in the jets of 1136--135 and
1150+497. Its nature (unresolved features vs. truly diffuse emission)
is unknown. Possible interpretations are discussed in \S~5.3.

\subsection{Spectral Energy Distributions} 

The Spectral Energy Distributions (SEDs) of the emission from the
various knots are the basis to investigate the origin of the emitted
radiation and the physical conditions in different regions of the
jets.  The SEDs of the knots are illustrated in Figures~9a-b, for
1136--135 and 1150+497 respectively. For clarity of presentation in
each panel the SEDs were shifted arbitrarily to allow comparison of
spectral variations along different locations in the jet.  Going from
bottom to top the SEDs run from the innermost to the outermost
detected knot, including the terminal hotspots.

While some of the SEDs were already presented in S02 and S04, the
deeper X-ray and optical and radio images have provided better
information about the jets morphologies, with a few additional knots
having been detected. Moreover, for the first time we are able to take
advantage of additional constraints provided simultaneously by the
optical and more accurate X-ray continuum slopes.

In the case of 1136--135, a spectral progression is clearly
present. The radio-to-optical continuum steepens systematically from
the innermost knot, $\alpha$, to the outermost one, E.  The radio and
optical continuum slopes also steepen, while the X-ray spectrum
flattens. For the two innermost knots, $\alpha$ and A, the optical
continuum appears to be flat. For knots B and C the optical spectra
are steep, indicating a break between the X-rays and the longer
wavelengths.

In 1150+497 there are less spectral variations. The radio-to-optical
continuum steepens, however, no clear trend is present for the
optical-to-X-ray continuum. Similarly, the radio and X-ray continua do
not change significantly from the inner to the outer regions of the jet,
with the X-rays maintaining a rather flat slope, inconsistent with an
extrapolation of the optical data.

In order to quantify the above trends we derived broad-band spectral
indices (Table~4). Using the 1~keV, 6250\AA, and 5 GHz fluxes in
Table~3, we derived the radio-to-X-ray, $\alpha_{rx}$, the
radio-to-optical, $\alpha_{ro}$, and the optical-to-X-ray,
$\alpha_{ox}$, spectral indices. Figures~10a-c show the plots of the
two-band energy indices versus the distance from the core in both
sources. In S04 we found that in all jets where multiwavelength
information is available for multiple knots there is a trend of
increasing radio-to-X-ray and radio-to-optical flux ratios with
increasing distance from the core. Figures~10 show that this trend is
confirmed for the jet of 1136--135, where both $\alpha_{rx}$ and
$\alpha_{ro}$ increase steeply with core distance. No trend is
observed for $\alpha_{ox}$ in this source. We recall that due to the
larger uncertainties of the optical fluxes and to the shorter
frequency range the latter index has larger errors.

For 1150+497 there are no demonstrable trends. A $\chi^2$ test gives
negligible probability of variation, i.e., in all three cases the run
of the energy index with core distance is consistent with a
constant. Inspection of Figure~8, however, shows that while the X-ray
emission fades systematically toward the end of the jet, the radio
increases abruptly at the position of the terminal hotspot. This is
likely related to a compression event that increases the synchrotron
emissivity through an increase of the magnetic field and electron
density (see \S~5.1). 

Inspection of the SEDs in Figures~9 shows that in all the knots the
optical points lie below the line connecting radio and X-rays,
indicating that a single power-law extending from the radio to the
X-rays can not reproduce the overall SEDs. Moreover, the different
slopes of the optical continuum suggest that the origin of the optical
emission can be different in different knots. Steep optical slopes
($\alpha_{opt}$ \gtsima 1) suggest that the optical emission belongs
to the high-energy tail of the synchrotron radio component, while
flatter slopes suggest that the optical belongs to the same spectral
component responsible for the X-ray emission. In some knots (notably
in knot A and B in 1136--135) the situation is quite unclear, since
the slope is poorly determined.  We refer to the next section for a
quantitative discussion of emission models for representative knots.

In S02, S04 we showed that the jet of 1150+497 and the outer knots of
the 1136--135 jet (C, D, E) had SEDs consistent with a model with two
emission components: synchrotron emission at longer wavelengths, and
Inverse Compton scattering of the CMB (IC/CMB) at higher energies,
implying a relativistic jet on kpc-Mpc scales. The inner knots of
1136--135 ($\alpha$ and A in this paper) were instead consistent with
synchrotron emission from radio to X-rays. The synchrotron
interpretation was suggested by the fact that available radio,
optical, and X-ray datapoints for $\alpha$ and A were consistent with
a single power-law. The new data presented here question this
interpretation, since the optical flux falls about a factor of 2 below
the radio-to-X-ray extrapolation. This result highlights the
importance of high quality data in studying multiwavelength emission
from these structures. 

\section{Modeling the multiwavelength jet emission} 

\subsection{Modeling the SEDs} 

We assume that the radio to X-ray emission from the knots of 1136--135
and 1150+497 is produced through synchrotron and IC/CMB scattering from a
relativistic electron population with a power law energy distribution. We
include the terminal knot in the analysis. Whether this correspond to the
traditional hotspot (Bridle et al. 1994) is an open question in this case. 
As shown in Tavecchio et al. (2005b), there is compelling evidence that,
as originally suggested by Georganopoulos \& Kazanas (2003), the emission
from these portions of the jet, usually considered at rest, is produced
by plasma still in motion with mildly relativistic speeds ($\Gamma \sim
2-3$). The advancement speed of the terminal shock front, however, can
be subrelativistic.

We refer to Tavecchio et al. (2000) and S02 for a full description of
the model used to reproduce the jet emission. Briefly, the emitting
regions are assumed here to be ellipsoidal, with volumes $V$
corresponding to the flux extraction regions used above (\S~3).  We
assume that the emitting region is in relativistic motion, with
Doppler factor $\delta$
and homogeneously filled by high-energy electrons, with a power-law
energy distribution $N(\gamma)=K\gamma^{-n}$ extending from $\gamma_{\rm
min}$ to $\gamma_{\rm max}$. Electrons radiate via the synchrotron and
IC/CMB mechanisms. The low-energy limit of the electron distribution
$\gamma_{\rm min}$ is well constrained by the condition that the
low-energy part of the IC/CMB component should not overproduce the
observed optical flux. To close the system of the model parameters we
assume that the energy densities of relativistic electrons and magnetic
field are in equipartition (protons are not included, i.e. $k=0$ in the
standard notation). This closely corresponds to the minimum energy
condition.

Specific models for different representative knots are discussed in
the following. The computed emission models are shown in Figures~11a-b
together with the observational data points.

For 1136--135 we consider first knots A and B which were detected in
all the three optical bands. The present data for knot A suggest a
moderately concave SED with the optical fluxes falling below the radio
to X-ray connection by a factor of about 2. Note that the radio flux
associated with this knot was revised upwards by a factor 5. At the
same time the optical slope is rather flat (though with large
uncertainty) indicating a connection of the optical to the X-ray
component.  The emission model we computed attributes the optical
emission mainly to the low energy tail of the IC/CMB, plus some
contribution from the high frequency extension of the synchrotron
component.  Previously this knot was interpreted as a pure synchrotron
knot.  For knot B the situation is similar to knot A but the optical
"deficiency" is larger and the optical slope steeper. Here the model
interpretes the optical fluxes as the synchrotron emission with a
minor contribution from the IC/CMB.  For comparison, we show in
Figure~11a the spectral model for knot C. Here the optical upper
limits separate the two components without ambiguity.
 
For 1150+497 we show in Figure~11b the spectral models computed for
knots B and C for which there is good optical information. In both
cases the optical slopes are steep suggesting a synchrotron origin of
the optical emission. Note that for knot B the flux level of the
optical emission falls close to a single power-law connecting the
radio and X-ray fluxes. The optical slope is crucial to determine
the model. 

Computing spectral models for all the knots with sufficient data, we
derived interesting parameters for different emission regions along
the jets, such as magnetic field $B$, the normalization of the
electron density $K$, the Doppler factor $\delta$. These are shown in
Figures~12a-b as a function of projected angular distance from the core. 
To determine the uncertainty associated to each derived quantity we
allow the input parameters to vary within a fixed range (5--25 for
$\gamma _{\rm min}$, $V-V/2$ for the volume), while for the radio flux,
the X-ray flux, and the radio slope we considered the errors reported in
Table 3.  We find that the uncertainty associated to the slope of
the radio spectrum (directly related to the index $n$ of the electron
energy distribution) largely dominates the final uncertainties,
reported by the the errorbars in Figures~12a-b. We believe that the
uncertainties estimated in this way are safely conservative, supporting
the reality of any variation af the parameters along the jet.

Inspection of Figures~12 shows a clear difference between the profiles
of the physical parameters of the two jets.  For 1136--135 a
systematic trend of all the parameters along the jet is apparent, in
the sense of increasing $K$ and $B$ and decreasing $\delta$.  For
1150+497 the profile is initially consistent with constant values
along the jet for a large part of its length, while only at the end
(close to the terminal HS) there is a significant increase of magnetic
field, electron density and a decrease of the Doppler
factor. Interestingly, the parameters estimated for the hotspot of
1136--135 are in good agreement with the trend followed by all the
previous knots, supporting the idea that the so called "hot spot" is
connected to the jet flow.

\subsection{Jet Deceleration?} 

One of the most striking differences between the two jets analyzed in
this work is related to the jets radio and X-rays
morphologies. Indeed, while the jet of 1150+497 presents a more or
less similar fading trend at all three wavelengths up to the terminal
hotspot, the jet of 1136--135 dramatically increases its radio
luminosity after knot B while the X-ray emission rapidly fades.  The
different brightness profiles directly translate in the different run
of the physical quantities from the SED modeling (Figures~12).  In
1136--135, the increasing radio brightness requires an increase of the
order of a factor 10 of the magnetic field and of the density of the
radiating electrons, bound together through the equipartition
condition. The decrease shown by the X-ray flux implies a decrease of
the Doppler factor, which reduces the IC/CMB emission.
Therefore, our analysis suggests that the observed morphology and
multiwavelength profiles of the 1136--135 jet could be the result of a
progressive deceleration of the emitting plasma flow after the region
corrisponding to knot B--C.

This conclusion, already indicated by our analysis of the jet of
3C~273 and of the multiwavelength properties of our jet sample
(Sambruna et al. 2001, S04), lends support to the model recently
proposed by Georganopoulos \& Kazanas (2004) for the deceleration of
the jet plasma on kpc scales.  In their analytical treatment the jet
is supposed to decelerate adiabatically, with an associated increase
through compression of the magnetic field and the particle
density. These effects lead to an increase of the synchrotron flux,
while the X-ray flux, produced through IC/CMB, decreases due to the
diminishing importance of beaming of the CMB photons.

A discussion of the the most important physical consequences of the
deceleration scenario will be presented in a companion paper (Paper
II). Here we just comment briefly on two of them.

In Paper II, we discuss entrainment of external gas as a possible
cause of the plasma deceleration in 1136--135 downstream of knot
B. Indeed, from the analysis of the ACIS images of the cores we find
evidence for the presence of circumnuclear hot gas in both sources on
a scale of several tens of kpc. Moreover, X-ray observations show that
low redshift ($z<1$) QSOs lie in environment typical of small groups
(Hardcastle \& Worrall 1999, Crawford \& Fabian 2003). In the case of
1150+497, deceleration seems to occur on a shorter (projected) length
and therefore is less apparent. Nevertheless, the possible scale
length in terms of distance from the core is of the same order.

A second interesting aspect of the deceleration model of Paper II is
that most of the ordered kinetic energy of the jet is converted during
the deceleration into internal random kinetic energy of protons,
increasing the total jet pressure. Only a small fraction of this energy
can be transmitted to the relativistic electrons, since the inferred
energy content in the relativistic electrons is limited to 10\% of the
dissipated energy. Comparing the radiative luminosity with the kinetic
energy flux the overall efficiency is very small, $10^{-4}$--$10^{-3}$.

\subsection{Inter-knot X-ray emission}

The detection of residual inter-knot X-ray emission is particularly
interesting in light of the deceleration scenario. We recall that the
spectral fits of the jet are consistent both with a hot (2--3 keV)
thermal component or with a non-thermal power law. 

The first possibility to be considered is that the emission is
produced by hot shocked plasma surrounding the jet (e.g., Komissarov
1994). In this case, we would expect values of the gas rotation 
measure from the radio in excess to the Galactic value. To constrain
the rotation measure in the 1136--135 jet, we made use of our 4.86 and
8.46 GHz \vla\ polarization images, along with our recalibration of an
archival 1.49 GHz \vla\ dataset published by Saikia et al. (1990). We
measure a rotation measure in the jet consistent with the integrated
galactic origin value of -26 $\pm$ 1 radians m$^{-2}$
(Simard-Normandin et al. 1981). Thus, we conclude that a thermal
origin of the diffuse inter-knot emission in the 1136--135 jet is
highly unlikely.

As anticipated in \S~4.3, the residual inter-knot X-ray emission could
be due to non-thermal emission from unresolved features associated to
the knot, spatially and/or physically. Among these possibilities is
that we are observing non-thermal emission produced by relativistic
electrons accelerated in a thin boundary layer around the jet. In the
model discussed by Stawarz \& Ostrowski (2002) the competition between
acceleration and radiation losses leads to a piled-up distribution,
whose peaked synchrotron emission can be quite hard (up to $\nu
^{1/3}$). The emission is most likely at X-rays: the peak energy is
given by $E=0.9 V_8$ keV, where $V_8$ is the characteristic velocity
of the magnetic turbulence expressed in units of $10^8$ cm s$^{-1}$,
expected to be of the order of the Alfven velocity, $V_A \sim 2\times
10^8$ cm s$^{-1}$ for typical parameters of the boundary layer. Thus,
the faint interknot X-ray emission observed in the sources of this
paper and others (S04) could be the synchrotron emission of hot
electrons piled-up in the jet boundary layer. 

\section{Discussion}

The previous discussion of the emission properties is based on the
synchro-IC/CMB model and a number of assumptions. In the following we
discuss some of the uncertainties and the theoretical problems
associated with this interpretation.

One necessary caveat regards the fluxes used to construct the SEDs and
the associated sizes of the emitting regions. We have already
discussed the arbitrary definition of ``knot'' in \S~4.3. Here we
comment on the physical implications of using a given finite region
size to extract the fluxes. Indeed, since images have different
angular resolution at different frequencies, we are forced to measure
fluxes using a quite large "aperture". However, the majority of the
optical features detected in the jets appear to be smaller than the
extraction regions. Therefore, we cannot exclude the possibility that
the size of the emission region is frequency-dependent. It is
conceivable that electrons are accelerated in a small volume and
subsequently diffuse from the acceleration site, radiatively cooling:
the high-frequency, optically emitting electrons will cool rapidly,
while radio and X-ray electrons, characterized by a small energy,
could diffuse into a larger volume. In this sense the use of a single
extraction region and a unique volume in the modeling could be
misleading.

Another important problem, related to the previous one, concerns the
nature of the emission regions (see also Stawarz et al. 2004). In the
modeling, the emitting region is assumed to be a "blob" of plasma, in
relativistic bulk motion. This is a reasonable assumption for many of
the observed knots, visible as isolated spots within the jet (e.g., A
in 1136--135). On the other hand, the morphology of 1136--135 after
knot B where it is difficult to define knot and interknot regions
suggests that a scenario considering emission from a continuous flow
may be more realistic (Paper II). We note that the possibility that
discrete knots mark portions of the jet characterized by larger
density and/or speed with respect the {\it average} jet could help to
overcome the difficulty regarding the large power requirement of
IC/CMB jets, pointed out by Dermer \& Atoyan (2004). In this case,
indeed, the {\it average} power of the jet, stored in the lobes, could
be significantly less than the ``istantaneous'' power inferred in
the knots. 

The above results were derived assuming that synchrotron+IC/CMB is the
mechanism responsible for the radio-to-X-ray emission in both jets.
Varius alternatives have been proposed in the past to explain the
peculiar shape of the SEDs of knots: proton synchrotron emission
(Aharonian 2002), synchrotron emission from a non power-law electron
distribution (resulting from cooling, Dermer \& Atoyan 2002, or from
acceleration effects, Stawarz et al. 2004), synchrotron emission from two
electron populations (Atoyan \& Dermer 2004). Among these models, those
invoking synchrotron emission from suitable electron distributions are
the most promising.

In the model proposed in Dermer \& Atoyan (2002), the underlying
electron population is characterized by an energy spectrum which
differs from the widely considered power-law form. The electron energy
spectrum derives from cooling of a power-law distribution extending to
large Lorentz factors (up to $\gamma \sim 10^8$). The cooling mechanism
is Inverse Compton on the CMB, which for high-energy electrons is
reduced because of the decline of the Klein-Nishina cross-section. As
a consequence the derived spectrum is characterized by a ``bump'' at
high energies.  The synchrotron spectrum produced by such a
distribution can reproduce the X-ray "excess" characterizing some of
the observed jets.  However, as discussed in Atoyan \& Dermer (2004),
the model can not explain the emission from knots with steep optical
spectra or knots in which the X-rays are much more luminous than the
optical, even assuming quite extreme values for the parameters. For
these cases the authors propose a two-component synchrotron
model. Except for the innermost knots of 1136--135, most of the knots
analized here exhibit a pronunced ``concave'' SED, for which the
Dermer \& Atoyan (2002) model can not be applied. The model proposed
by Stawarz et al. (2004) encounters a similar difficulty.

Another uncertainty of the model is associated to the use of the
equipartition condition in deriving the physical parameters of the
knots. Stawarz et al. (2005) recently showed that, for the specific case
of knot A in the jet of M87, the present upper limits on the SSC emission
at high energy suggest a deviation from equipartition condition, with the
magnetic field dominating the total energy. A possibility to explain such
large magnetic field is that it is amplified by dynamo effects induced by
turbulence. However, the present knowledge of this problem is still
rather poor.  Note that, relaxing the equipartition hypothesis, allowing
for a dominance of the magnetic field (electrons), leads to larger
(lower) values of the beaming factor. Moreover, the carried power will
increase out of equipartition. 

Clearly, several uncertainties and open questions remain in our
understanding of the multiwavelength emission of large-scale
jets. However, we believe that a simplified approach based on
``individual'' emission regions and equipartition is useful, as it
leads to an unambiguous derivation of the physical conditions which
can then be confronted with independent constraints.

\section{Summary}

We have presented deep \chandra\ ACIS and multi-color \hst\ ACS
observations of the jets of two powerful quasars, selected from our
previous X-ray and optical survey (S04). The exposures and choice of
filters for the ACS were optimized to obtain more detailed jet
morphologies as well as X-ray and optical spectra for individual
bright knots. The following results were obtained: 

\begin{itemize} 

\item All the jet features previously detected at X-rays and optical
are confirmed. A few faint knots were detected for the first time,
illustrating the importance of deeper exposures to gain a better
knowledge of the jet structure. 

\item The jet profiles at the various wavelengths show in general
good correspondence among the knots. A qualitatively different jet
profile was observed in the two sources. In 1136--135, the jet X-ray
emission fades after the most prominent knot B while the radio
emission simultaneously increases. In 1150+497, the emission at all
wavelengths steadily decreases, with the exception of the radio flux
which increases again in the terminal jet region. 

\item The optical data indicate steep optical spectra in agreement 
with the expectations from the IC/CMB model, except for the first two
knots of 1136--135.

\item Large spectral variations are observed along the jet of
1136--135. Here, $\alpha_{rx}$ and $\alpha_{ro}$ steepen with
distance from the core, and $\alpha_{ro}$ steepens with increasing
radio luminosity of the knots. 

\item The multiwavelength morphology and SED variations in 1136--135 are
consistent with the idea that the plasma suffers substantial deceleration
on kpc-scales, as previously suspected for 3C~273 and other jets. A
physical model is discussed in Paper II.

\item The X-ray emission profile of the jets consists of a non-zero
baseline over which more or less defined peaks are superposed. The
origin of the non-zero inter-knot X-ray emisison is unclear, but a
strong possibility consistent with the present data is non-thermal
emission from unresolved features, such as oblique shocks or hot
electrons piled-up in the jet boundary layer (Stawarz \& Ostrowski
2002). 

\end{itemize}

\acknowledgements

We are grateful to the anonymous referee, who provided excellent and
constructive criticism. This project is funded by NASA grant
HST-GO4-5111A, which is operated by AURA, Inc., and GO2-3195C and
HST-GO-09122.08-A (CCC, JFCW) from the Smithsonian Observatory. RMS
gratefully acknowledges support from an NSF CAREER award and from the
Clare Boothe Luce Program of the Henry Luce Foundation while at George
mason University. MG and RMS acknowledge funds from NASA LTSA grant
NAG5--10708. Radio astronomy at Brandeis is supported by the NSF. The
National Radio Astronomy Observatory is operated by Associated
Universities Inc. under a cooperative agreement with the National
Science Foundation. \merlin\ is a National Facility operated by the
University of Manchester at Jodrell Bank Observatory on behalf of
PPARC.


\vskip 1 cm

\begin{center}
\begin{tabular}{llcrlrrrr}
\multicolumn{9}{l}{{\bf Table 1: The Targets}} \\
\multicolumn{9}{l}{   } \\ \hline
& & & & & & & & \\
Source & $z$ & Gal N$_H$ & logP$^{5~GHz}_{core}$ & logP$^{5~GHz}_{jet}$ & $R_i$ & m$_V$ & log L$_{BLR}$ & log L$_{[OIII]}$ \\
& & & & & \\
(1) & (2) & (3) & (4) & (5) & (6) & (7) & (8) & (9) \\
& & & & & & & & \\ \hline
& & & & & & & & \\
1136--135& 0.554 & 3.5& 33.73 & 33.69 & 0.30 & 16.1 & 45.37 & 43.86 \\
         &       &       &      &   &    & & &  \\
& & & & & & & & \\
1150+497 & 0.334 & 2.0 & 34.53 & 34.95 & 2.3 & 17.1 & 45.80 & 43.75  \\
& & & & & & & & \\ \hline

\end{tabular}
\end{center}

\noindent
{\bf Explanation of Columns:} 1=Source IAU name; 2=Redshift;
3=Galactic column density in $10^{20}$ \nh; 4=Log of the core power at
5 GHz (in erg s$^{-1}$ Hz$^{-1}$); 5=Log of the jet power at 5 GHz (in
erg s$^{-1}$ Hz$^{-1}$); 6=Ratio of core to extended (total-core)
radio power at 5 GHz, corrected for the redshift (observed value times
(1+z)); 7=Core optical V magnitude; 
8=Total luminosity of the BLR (Cao \& Jiang 1999); 
9=Luminosity of the [OIII] line (Cao \& Jiang 1999). 


\clearpage

\scriptsize 
\begin{center}
\begin{tabular}{lcccrcc}
\multicolumn{7}{l}{{\bf Table 2: X-ray Observations}} \\
\multicolumn{7}{l}{   } \\ \hline
Knot & Dist & FWHM & $\Delta$ FWHM & Counts & $\Gamma_X$ & Opt? \\
& & & & &  & \\
& (\arcsec) & (\arcsec) & (\arcsec) &  &  & \\ 
& & & & &  & \\ \hline
\multicolumn{7}{c}{\bf 1136--135} \\ \hline 
& & & & & & \\
$\alpha$ & 2.7 &0.75 & 0.45  &  98$\pm$11 & 1.9$^{+0.4}_{-0.4}$ & Y \\
A        & 4.6 &0.85 & 0.53    & 91$\pm$11 & 2.1$^{+0.3}_{-0.6}$ & Y \\
B        & 6.5 &0.6 &  0.0    & 202$\pm$14 & 2.1$^{+0.2}_{-0.3}$ & Y \\
C        & 7.7 &0.6 &  0.0    & 90$\pm$11 & 1.5$^{+0.3}_{-0.2}$ & N \\
D        & 8.6 & $\cdots$ & $\cdots$&  28$\pm$6  & 1.5$^{+0.5}_{-0.5}$ & Y \\
E        & 9.3 & $\cdots$ &  $\cdots$ & 21$\pm$6  & 2.3$^{+0.6}_{-0.5}$ & N \\
HS  & 10.3& $\cdots$ &  $\cdots$ & 12$\pm$5  & 1.7$^{+0.9}_{-0.7}$ & Y  \\
& & & & & & \\ \hline
\multicolumn{7}{c}{\bf 1150+497} \\ \hline 
& & & & & & \\
B & 2.2 &0.76 &  0.36  & 250$\pm$16 & 1.7$^{+0.2}_{-0.2}$ & Y \\
C & 2.6 &0.91 &  0.53  & 80$\pm$10 & 1.5$^{+0.3}_{-0.3}$ & Y \\ 
D & 3.3 & $\cdots$ & $\cdots$ &  28$\pm$6  & 1.7$^{+0.5}_{-0.5}$ & Y \\ 
E & 4.3 &0.71 & 0.36  & 84$\pm$10 & 1.7$^{+0.3}_{-0.2}$ & Y \\ 
F & 5.4 & $\cdots$ & $\cdots$ & 35$\pm$7  & 1.5$^{+0.5}_{-0.5}$  & N \\
G & 6.7 &$\cdots$ &  $\cdots$ & 31$\pm$7  & 1.6$^{+0.6}_{-0.4}$  & N \\
H & 8.0 &$\cdots$ &  $\cdots$ & 33$\pm$7  & 1.7$^{+0.3}_{-0.6}$  & Y \\
IJ & 8.4 & $\cdots$ & $\cdots$ & 19$\pm$5  & 2.1$^{+0.6}_{-0.6}$  & Y \\
\hline 
& & & & & & \\
\end{tabular}
\end{center}

\noindent
{\bf Explanation of Columns:} 1=Source knot. The designation HS is for
Hot Spot; 2=Distance from core; 3=FWHM of the knot in 0.3-8 keV in the
direction orthogonal to the jet's axis;
4=Difference in quadrature between the FWHM in column 3 and the core's
FWHM; 5=Source net counts in 0.3--8 keV; 6=Photon Index from a power
law plus Galactic N$_H$ fit; 7=Optical counterpart: Y=Yes, N=No.


\vskip 1 cm

\scriptsize 
\begin{center}
\begin{tabular}{lrrrrrrrrrrrl}
\multicolumn{13}{l}{{\bf Table 3: Multiwavelength Fluxes of Jets Knots}} \\
\multicolumn{13}{l}{   }\\
 \hline
\hline
\multicolumn{13}{c}{\bf 1136--135} \\ 
\hline 
& & & & & & & & & & & \\ 
Knot && F$_{1~keV}$ && F$_{8140~\AA}$ & F$_{6250~\AA}$ & F$_{4750~\AA}$ 
        & $\alpha_{opt}$  && F$_{4.9~GHz}$ & F$_{8.5~GHz}$ & F$_{22.5~GHz}$ & $\alpha_{rad}$\\ 
& & & & & & & & & & & \\ \hline 
$\alpha$ && 1.9$\pm$0.2 && 0.17$\pm$0.05 & 0.15$\pm$0.03 & 0.13$\pm$0.04  
   &0.51 $\pm$ 0.27 && 4.4$^{\dagger}$ & 3.0 $\pm$ 0.3 & 1.5$^{\dagger}$ & 0.75 $\pm$ 0.10 \\
A && 1.7$\pm$0.2 && 0.23$\pm$0.04 & 0.16$\pm$0.02 & 0.17$\pm$0.02 
   &0.27 $\pm$ 0.17 && 5.4$^{\dagger}$ & 3.8$\pm$0.4 & 2.0$^{\dagger}$  & 0.67 $\pm$ 0.11 \\ 
B && 3.5$\pm$0.2 && 0.33$\pm$0.06 & 0.33$\pm$0.02 & 0.22$\pm$0.02 &1.17 $\pm$ 0.19
 && 10.9 $\pm$ 1.6 & 9.3 $\pm$ 0.9 & 3.2 $\pm$ 0.6  & 0.81$\pm$0.13\\
C && 1.8$\pm$0.2 && $<$0.09 & $<$ 0.09 & $<$0.07 & $\cdots$ && 
           31.5 $\pm$ 3.2 & 20.6 $\pm$ 2.1 & 9.2 $\pm$ 1.8  & 0.66$\pm$0.09\\ 
D && 1.0$\pm$0.2 && $<$0.22 & 0.20$\pm$0.03 & $<$0.08 & $\cdots$ && 
           48.6 $\pm$ 4.9 & 29.5 $\pm$ 3.0 & 11.9 $\pm$ 2.4  & 0.71$\pm$0.08\\
E && 0.7$\pm$0.2 && $<$0.15 & $<$0.05 & $<$0.08 & $\cdots$ &&  
           111.5 $\pm$ 11.2 & 66.1 $\pm$ 6.6 & 26.4 $\pm$ 5.3  & 0.82$\pm$0.09\\
HS && $<$0.6 && 0.24$\pm$0.03 & 0.15$\pm$0.03 & $<$0.08 & 1.85 && 
         199.9 $\pm$ 20.0 & 119.0 $\pm$ 11.9 & 43.9 $\pm$ 8.8  & 0.85$\pm$0.08\\ 
& & & & & & & & & \\ \hline
\multicolumn{13}{c}{\bf 1150+497} \\ \hline 
& & & & & & & & & & \\
Knot && F$_{1~keV}$ && F$_{8140~\AA}$ & F$_{6250~\AA}$ & F$_{4750~\AA}$ 
        & $\alpha_{opt}$  && F$_{1.7~GHz}$ & F$_{4.9~GHz}$ & F$_{22.5~GHz}$ & $\alpha_{rad}$ \\ 
& & & & & & & & & & & \\ \hline 
B && 7.6$\pm$0.5 && 1.13$\pm$0.15 & 0.99$\pm$0.07 & 0.64$\pm$0.04 
   &1.31 $\pm$ 0.50 &&35.7$\pm$5.4 & 17.3$\pm$1.7 & 5.4$\pm$1.1& 0.72$\pm$0.09 \\ 
C && 2.9$\pm$0.4 && 0.27$\pm$0.04 & 0.21$\pm$0.02 & 0.13$\pm$0.02 &
  1.49 $\pm$ 0.17&&26.1$\pm$3.9 & 13.0$\pm$2.0 & 4.0$\pm$1.0 & 0.71$\pm$0.10\\
D && 1.3$\pm$0.3 && 0.10$\pm$0.02 & 0.09$\pm$0.02 & $<$0.10 &0.43 && 
        12.3$\pm$1.8 & 6.5$\pm$1.0  & 2.0$\pm$0.5 & 0.68$\pm$0.10\\
E && 1.7$\pm$0.2 && 0.13$\pm$0.03 & 0.10$\pm$0.01 & $<$0.08 & 1.18&& 
        26.5$\pm$4.0 & 14.8$\pm$1.5 & 4.6$\pm$0.9  & 0.67$\pm$0.08\\
F && 1.0$\pm$0.2 && $<$0.10 & $<$0.07 & $<$0.03 & $\cdots$ && 
        11.5$\pm$2.3 & 9.1$\pm$1.4  & 2.3$\pm$0.7  & 0.59$\pm$0.11\\
G && 0.8$\pm$0.2 && $<$ 0.04  & $<$0.05 & $<$ 0.01 & $\cdots$ &&  
        8.1$\pm$1.6 & 4.4$\pm$0.7  & 0.9$\pm$0.3 & 0.81$\pm$0.11\\
H && 1.0$\pm$0.2 && $<$ 0.04 &$<$  0.05 & $<$ 0.04 & $\cdots$ && 
        24.4$\pm$3.7 & 12.9$\pm$1.9 & 3.7$\pm$0.7 & 0.72$\pm$0.09\\
IJ && 0.6$\pm$0.1 && 0.11$\pm$0.02 & 0.10$\pm$0.02 & 0.05$\pm$0.01 
&1.64 $\pm$ 0.09 && 64.3$\pm$9.6 & 28.5$\pm$4.3 & 7.6$\pm$1.9 & 0.81$\pm$0.10\\
\hline 
& & & & & & & & & & \\
\end{tabular}
\end{center}

\noindent
{\bf Notes:} $^{\dagger}$=Extrapolated using the slope in column 13 (see text). 
{\bf Explanation of Columns:} 1=Source knot. The designation HS is for
Hot Spot; 2=X-ray flux density at 1 keV in $\mu$Jy, corrected for absorption; 
3,4,5=Optical flux densities at the indicated wavelengths 
in $\mu$Jy, corrected for absorption. The upper limits are 3$\sigma$; 
6=Optical energy index from a linear interpolation to the optical fluxes (see text); 
7,8,9=Radio flux densities at the indicated frequencies in mJy;
10=Radio energy index.


\vskip 1 cm

\scriptsize 
\begin{center}
\begin{tabular}{lccc}
\multicolumn{4}{l}{{\bf Table 4: Broad-band spectral indices of Jets}} \\
\multicolumn{4}{l}{   } \\ \hline
Knot & $\alpha_{ro}$ & $\alpha_{ox}$ &  $\alpha_{rx}$ \\
& & & \\
(1) & (2) & (3) & (4)\\
& & & \\ \hline
\multicolumn{4}{c}{\bf 1136--135} \\ \hline 
& & & \\
$\alpha$ & 0.90$\pm$0.10 & 0.70$\pm$0.10 & 0.83$\pm$0.06\\
A        & 0.91$\pm$0.07 & 0.73$\pm$0.07 & 0.85$\pm$0.07 \\
B        & 0.91$\pm$0.07 & 0.73$\pm$0.04 & 0.84$\pm$0.07 \\
C        &  $>$1.11     & $<$ 0.63      & 0.94$\pm$0.07 \\
D        & 1.08$\pm$0.08 & 0.85$\pm$0.11 & 1.00$\pm$0.10 \\
E        &  $>$ 1.28 & $<$ 0.68 & 1.07$\pm$0.13\\
HS       & 1.23$\pm$0.10 & $>$ 0.88 & $>$ 1.11 \\
& & & \\ \hline
\multicolumn{4}{c}{\bf 1150+497} \\ \hline 
& & &  \\
B & 0.85$\pm$0.05 & 0.78$\pm$0.04 & 0.83$\pm$0.05 \\ 
C & 0.96$\pm$0.08 & 0.69$\pm$0.07 & 0.87$\pm$0.09 \\
D & 0.98$\pm$0.12 & 0.68$\pm$0.14 & 0.87$\pm$0.12 \\
E & 1.04$\pm$0.06 & 0.65$\pm$0.07 & 0.90$\pm$0.07 \\
F & $>$1.03       & $<$ 0.68      & 0.91$\pm$0.11 \\
G &  $>$0.99 & $<$ 0.66 & 0.88$\pm$0.13 \\
H &  $>$1.09 &$<$ 0.63 & 0.92$\pm$0.11 \\
IJ& 1.10$\pm$0.11 & 0.82$\pm$0.11 & 1.00$\pm$0.10 \\
& & & \\ \hline 
\end{tabular}
\end{center}

\noindent
{\bf Explanation of Columns:} 1=Source knot. The designation HS is for
HotSpot; 2=Radio-to-optical index; 3=Optical-to-X-ray
index; 4=X-ray-to-radio index. 


\newpage 

\begin{figure}
\centerline{\includegraphics[height=7.0in,width=7.5in]{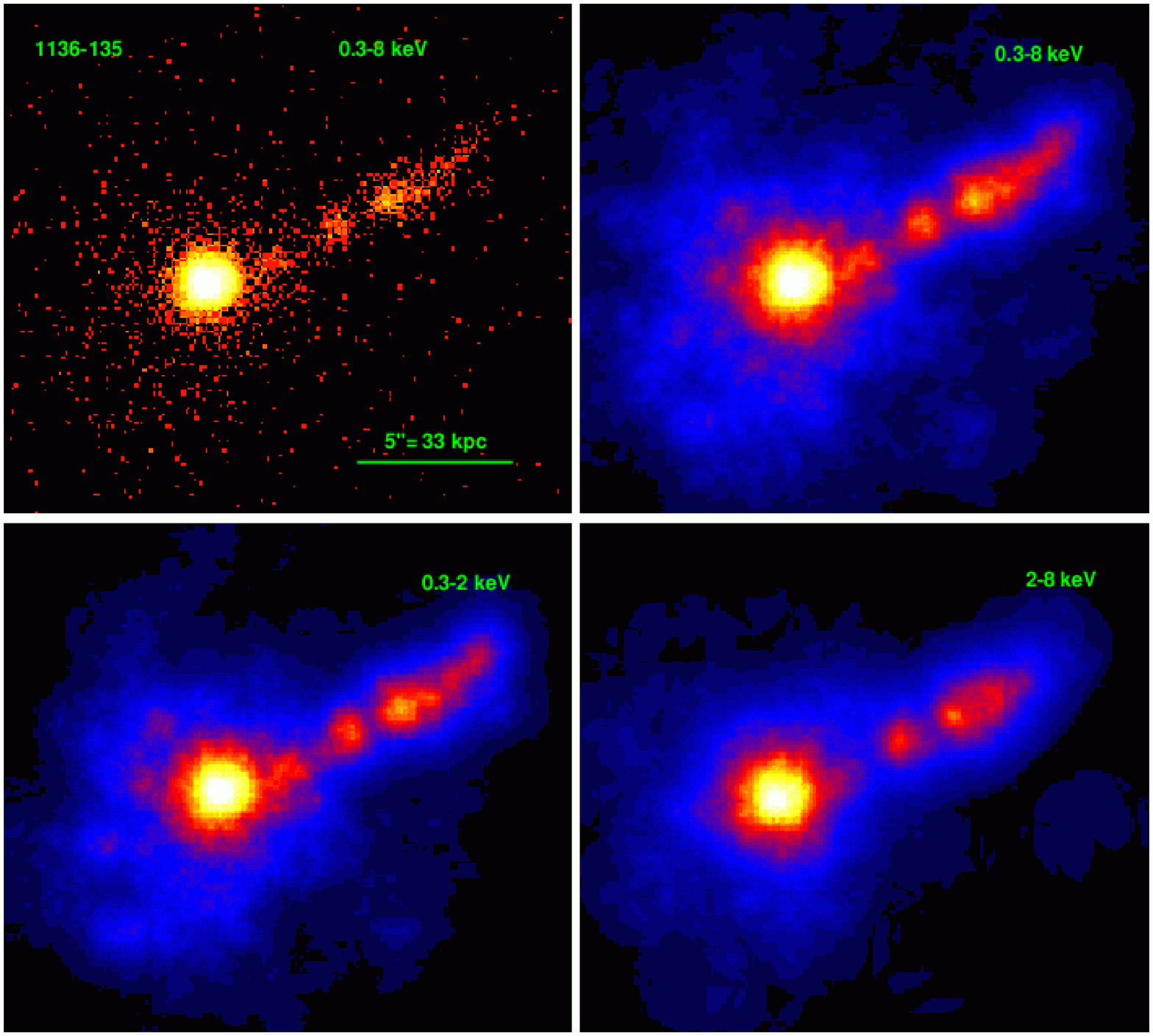}}
\caption{X-ray images of the jet of 1136--135 from a 77.4~ks \chandra\
ACIS-S exposure. Top, Left: raw ACIS image in total band. Top, Right
to Bottom, Right: smoothed ACIS images in total, soft, and hard band,
respectively. In the smoothed images, the pixel size is
0.1\arcsec. North is up and East to the left.}
\end{figure}

\newpage 

\begin{figure}
\centerline{\includegraphics[height=7.0in,width=7.5in]{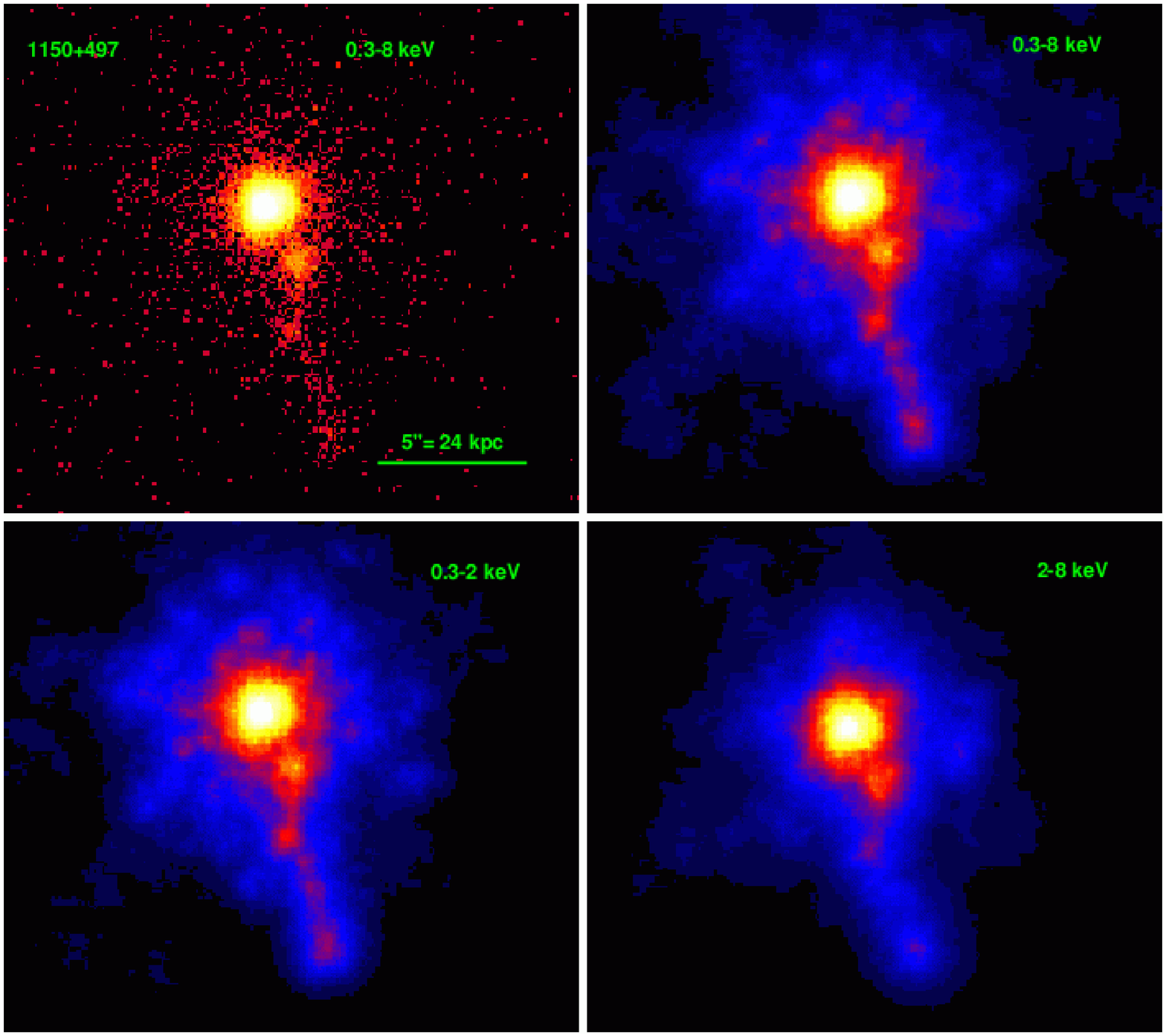}}
\caption{X-ray images of the jet of 1150+497 from a 68~ks \chandra\
ACIS-S exposure. Top, Left: raw ACIS image in total band. Top, Right
to Bottom, Right: smoothed ACIS images in total, soft, and hard band,
respectively. In the smoothed images, the pixel size is
0.1\arcsec. North is up and East to the left.} 
\end{figure}
\newpage 


\begin{figure}
\caption{Multiwavelength imaging of the 1136--135 jet. The color
images are the smoothed ACIS image in 0.3--8 keV (top), the ACS image
in the F625W filter (center), and the 22~GHz image (bottom). The 8.46
GHz radio contours are overlaid to all images, smoothed with a
resolution comparable to the other wavelengths (0.1\arcsec). The beam
size is 0.253\arcsec x 0.157\arcsec\ at PA = 78.6\deg\ at 22~GHz and
0.5\arcsec\ at 8.46~GHz. The radio contours range from --0.005 and
0.05 in logarithmic scale.  The radio knots are labeled. North is up
and East to the left.}
\end{figure}
\newpage

\begin{figure}
\caption{Multiwavelength imaging of the 1150+497 jet. The color
images are the smoothed ACIS image in 0.3--8 keV (top), the ACS image
in the F625W filter (center), and the 22~GHz image (bottom). The 4.9
GHz radio contours are overlaid to all images, smoothed with a
resolution comparable to the other wavelengths (0.1\arcsec). The beam
size is 0.315\arcsec x 0.289\arcsec\ at PA = 86.3\deg\ at 22~GHz and 
0.5\arcsec\ at 8.46~GHz. The radio knots are labeled. North is up and 
East to the left.}
\end{figure}
\newpage 


\begin{figure}
{\includegraphics[height=7.0in,width=6.5in]{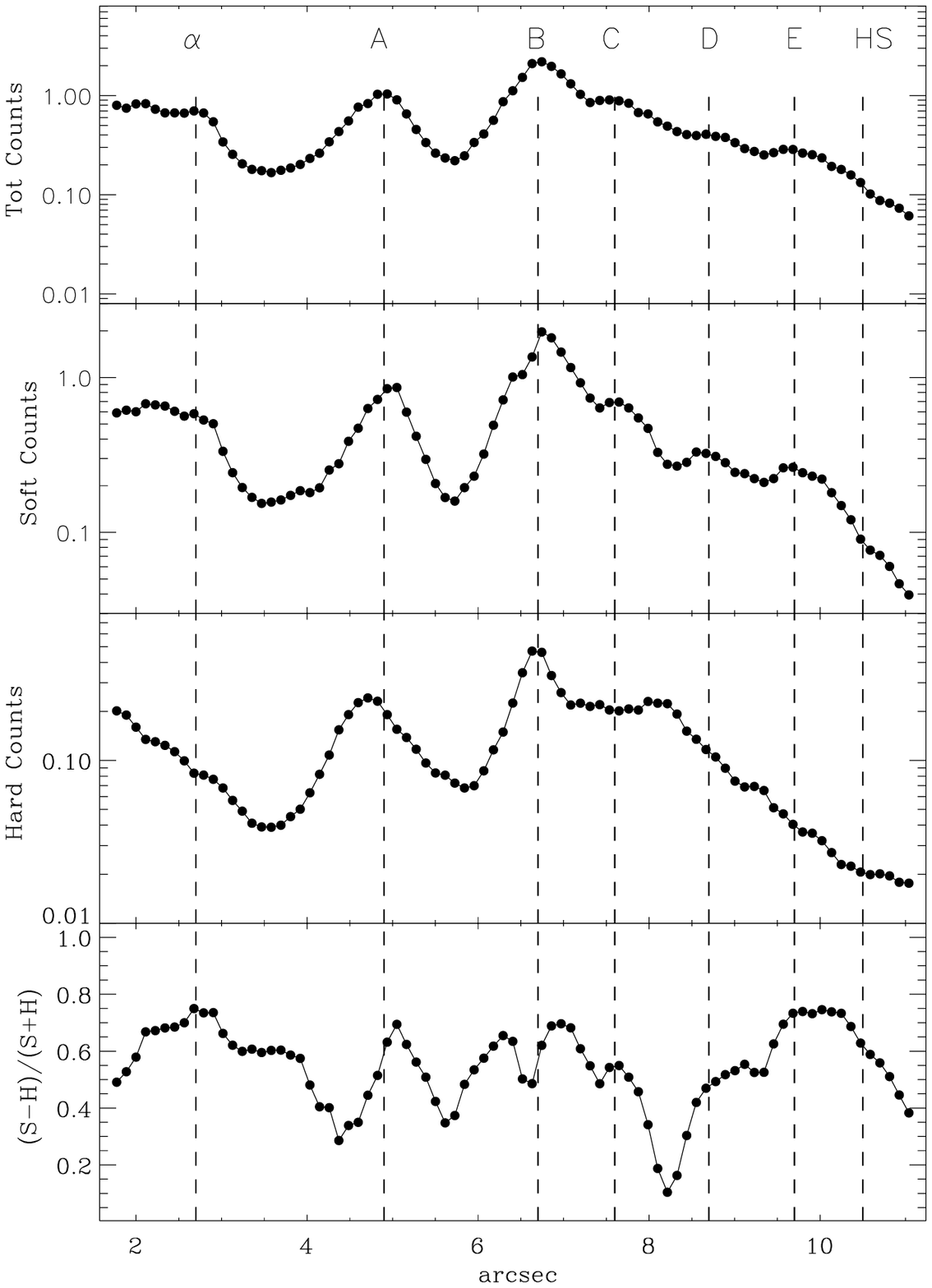}}
\vskip 0.7 cm
\caption{X-ray profiles of the 1136--135 jet at various
energies (top three panels) and run of the X-ray softness ratios along
the jet (bottom panel). The vertical axis represents the average
counts/pixel of the linear profile measured using a constant
box. Since the thickness of the box is ~15 pixels (corresponding
1.5"), all values should be multiplied by 15 to have the total number
of counts along each strip of the jet. The vertical dashed lines mark
the position of the radio/X-ray features (Table~2). Note the
qualitatively different behavior of the jet before and after knot
B. Before this knot, the jet consists of isolated knots while
downstream of knot B the flow is almost continuous.}
\end{figure}
\newpage 
\begin{figure}
{\includegraphics[height=7.0in,width=6.5in]{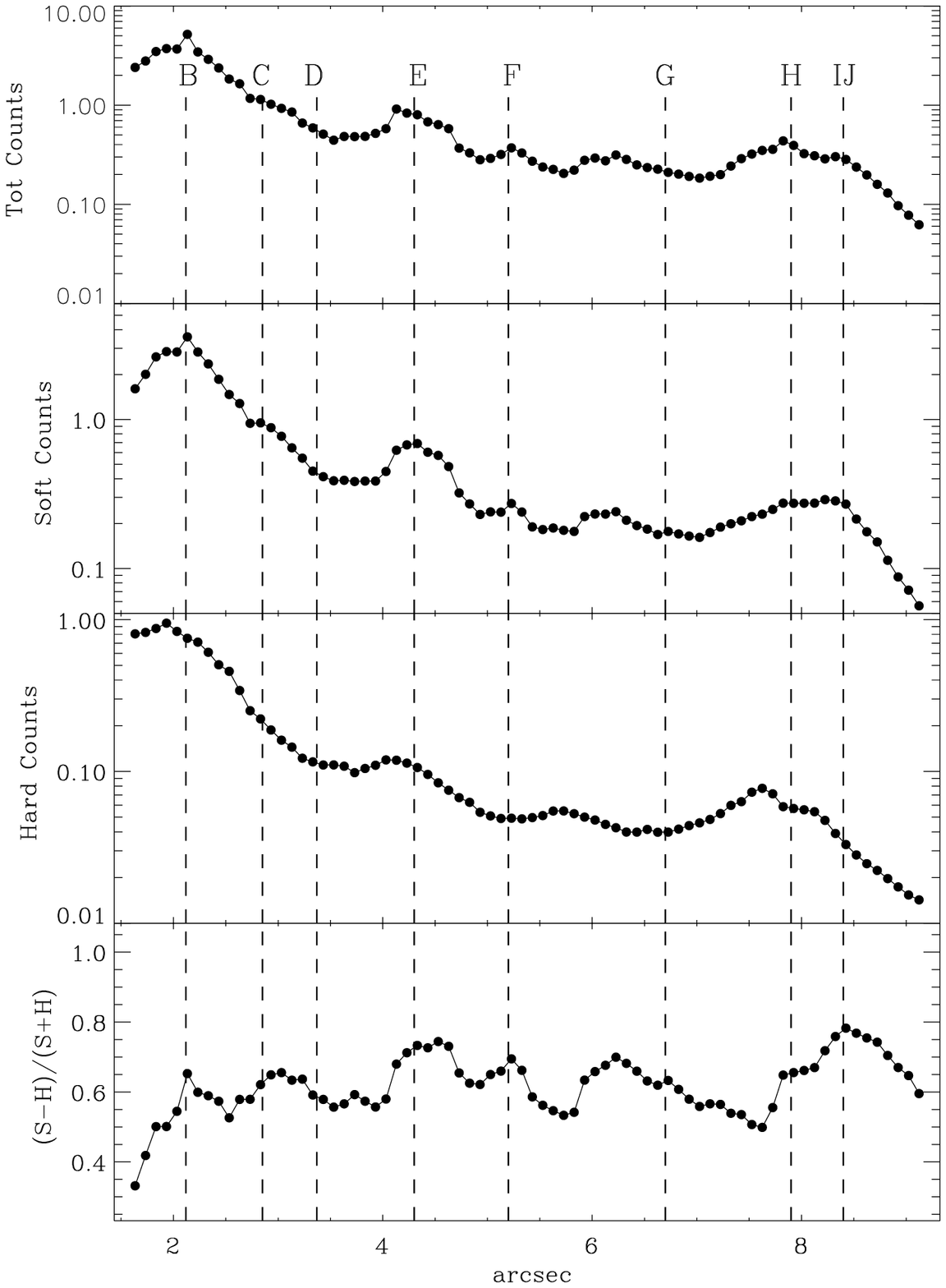}}
\vskip 0.7 cm
\caption{X-ray profiles of the 1150+497 jet at various
energies (top three panels) and run of the X-ray softness ratios along
the jet (bottom panel). The vertical axis represents the average
counts/pixel of the linear profile measured using a constant
box. Since the thickness of the box is ~15 pixels (corresponding
1.5"), all values should be multiplied by 15 to have the total number
of counts along each strip of the jet. The vertical dashed lines mark
the position of the radio/X-ray features (Table~2). 
Because of the wiggling structure of the jet some of the
knots detected in the image are not easily discernible here. The
multiwavelength emission decreases steadily toward the end of the
jet.}
\end{figure}
\newpage 

\begin{figure}
{\includegraphics[height=6.5in,width=6.5in]{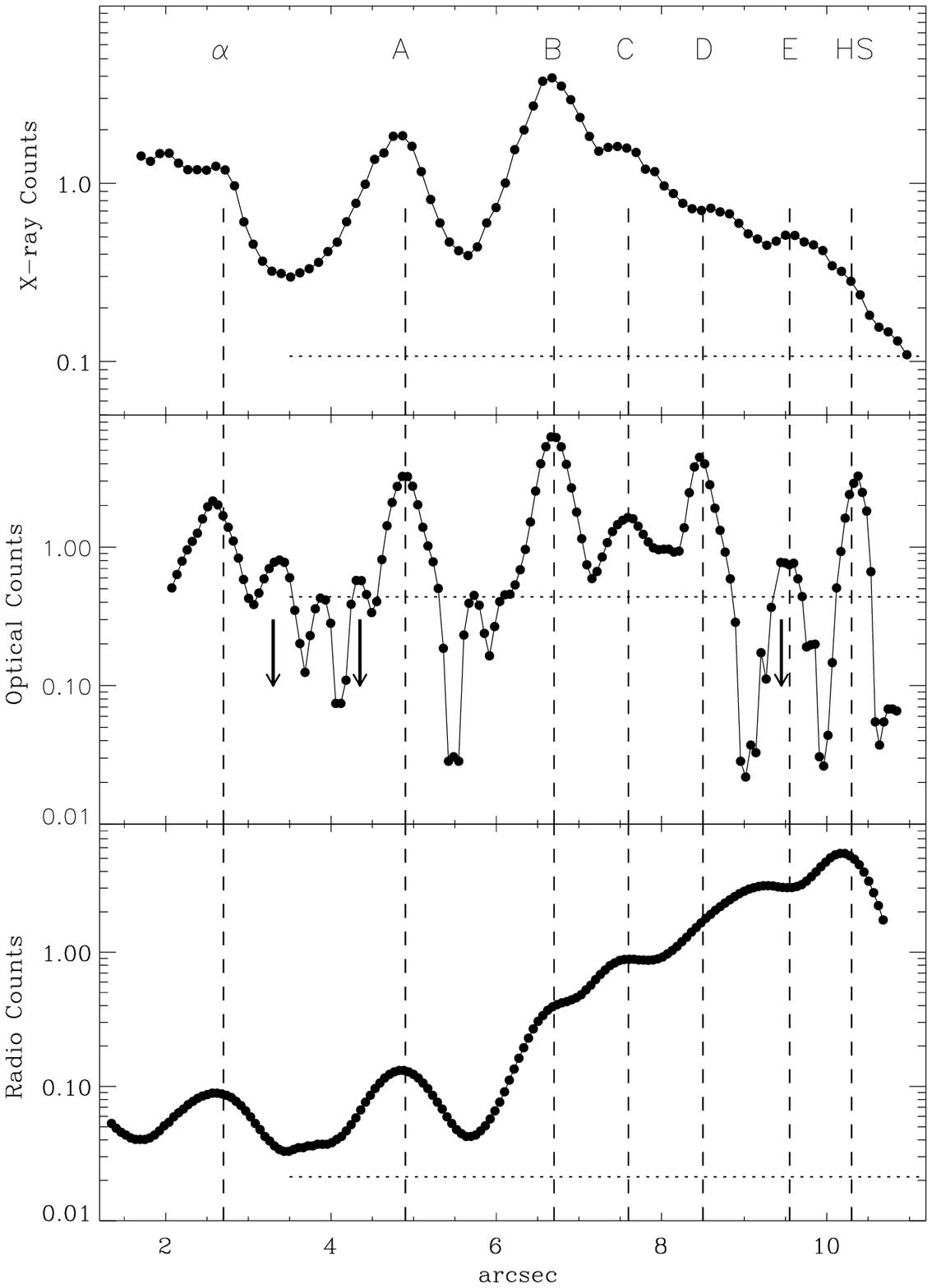}}
\vskip 0.7 cm
\caption{Multiwavelength profiles of the 1136--135 jet (X-rays,
top; optical, middle; radio, bottom). The horizontal dotted lines mark
the background levels; this is an average in the case of the optical
profile, where the background varies with pixel location. The profiles
were normalized to the average value of the counts/pixel (0.561 for
the X-rays, 0.00457 for the optical, and 0.00943 for the radio). The arrows
in the optical profile mark those knots which were not detected
because of higher local background. There is in general a good
correspondence between the positions of the the knots at the three
wavelengths. Note the different behavior of the jet at X-rays and
radio after knot B; the X-ray emission fades while the radio increases
toward the end of the jet.}
\end{figure}
\newpage 
\begin{figure}
{\includegraphics[height=6.5in,width=6.5in]{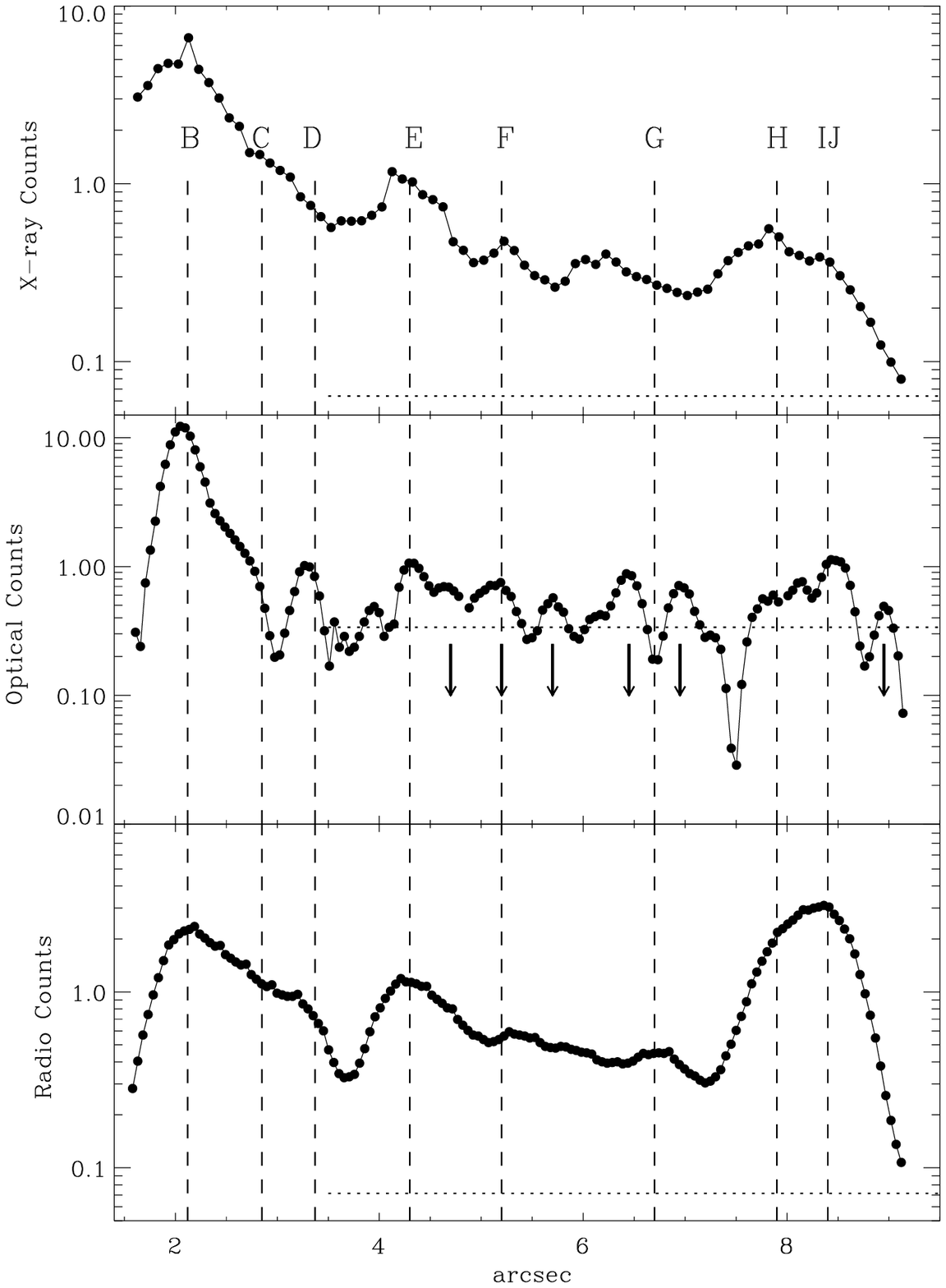}}
\vskip 0.7 cm
\caption{Multiwavelength profiles of the 1150+497 jet (X-rays,
top; optical, middle; radio, bottom). The horizontal dotted lines mark
the background levels; this is an average in the case of the optical
profile, where the background varies with pixel location. The profiles
were normalized to the average value of the counts/pixel (0.782 for
the X-rays, 0.00593 for the optical, and 0.00280 for the radio). The arrows
in the optical profile mark those knots which were not detected
because of higher local background. There is a general good
correspondence between the positions of the the knots at the three
wavelengths. Note the sudden increase of the radio flux at the end of
the jet.}
\end{figure}
\newpage 


\begin{figure}
\noindent
{\includegraphics[height=11cm,width=8cm]{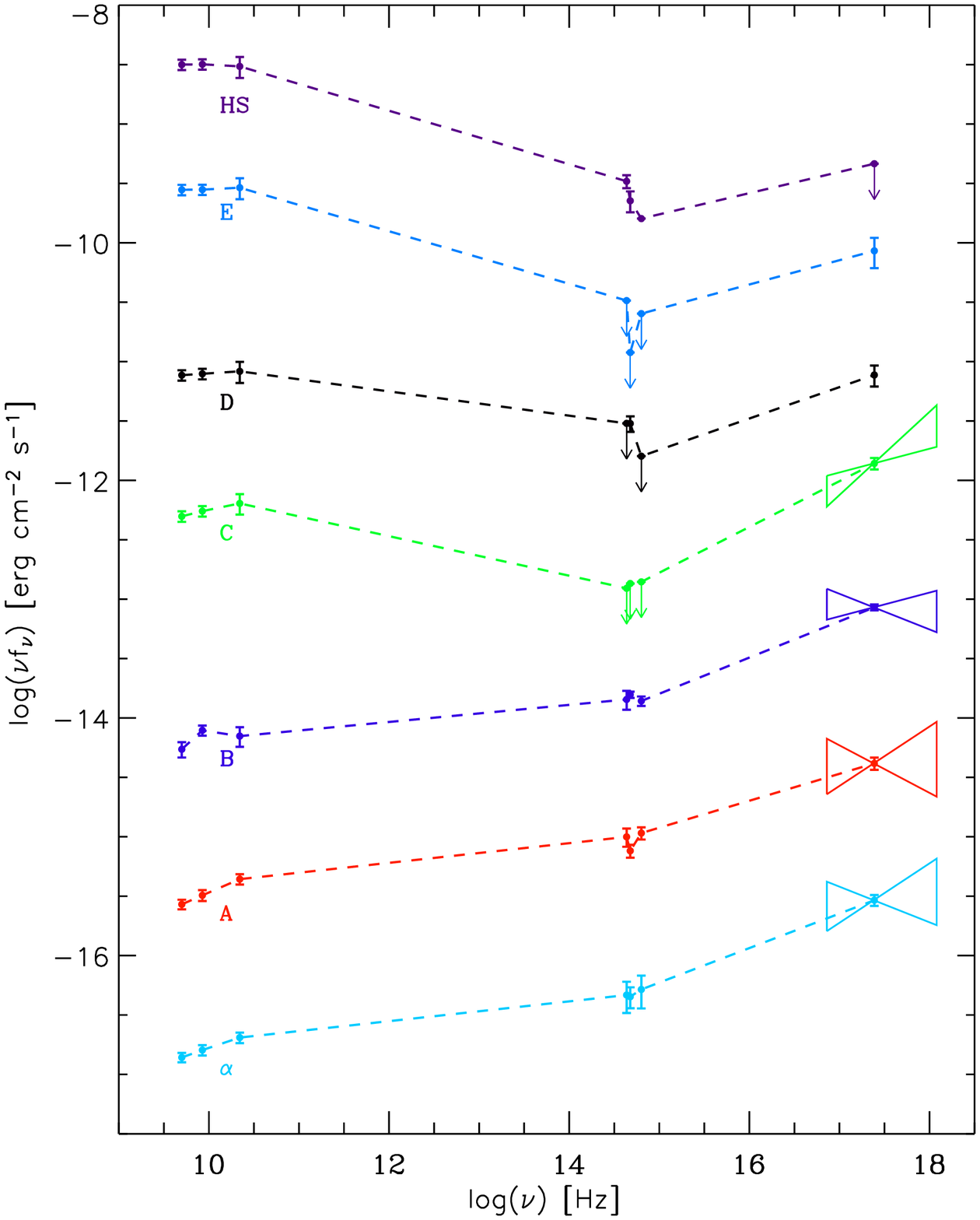}}{\includegraphics[height=11cm,width=8cm]{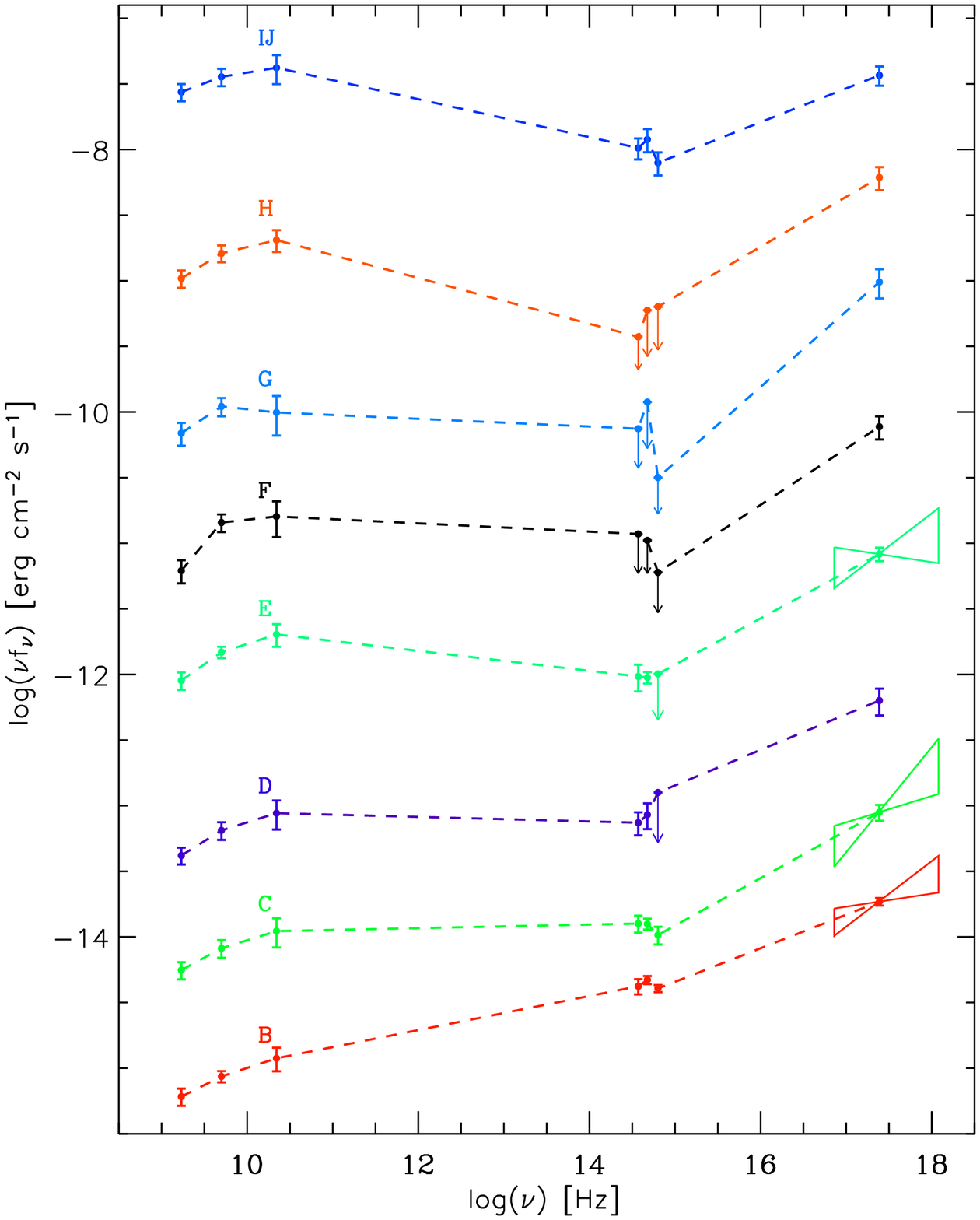}}
\caption{Spectral Energy Distributions (SEDs) of the jets of 1136--135
(left) and 1150+497 (right). In each panel, going from bottom to top 
the SEDs are plotted from the inner to the outer jet, with arbitrary
shifts for clarity. Spectral changes are observed in the jet of
1136--135, while the SED shape of 1150+497 remains fairly constant.}
\end{figure}

\newpage 


\begin{figure}
\noindent
{\includegraphics[height=9cm]{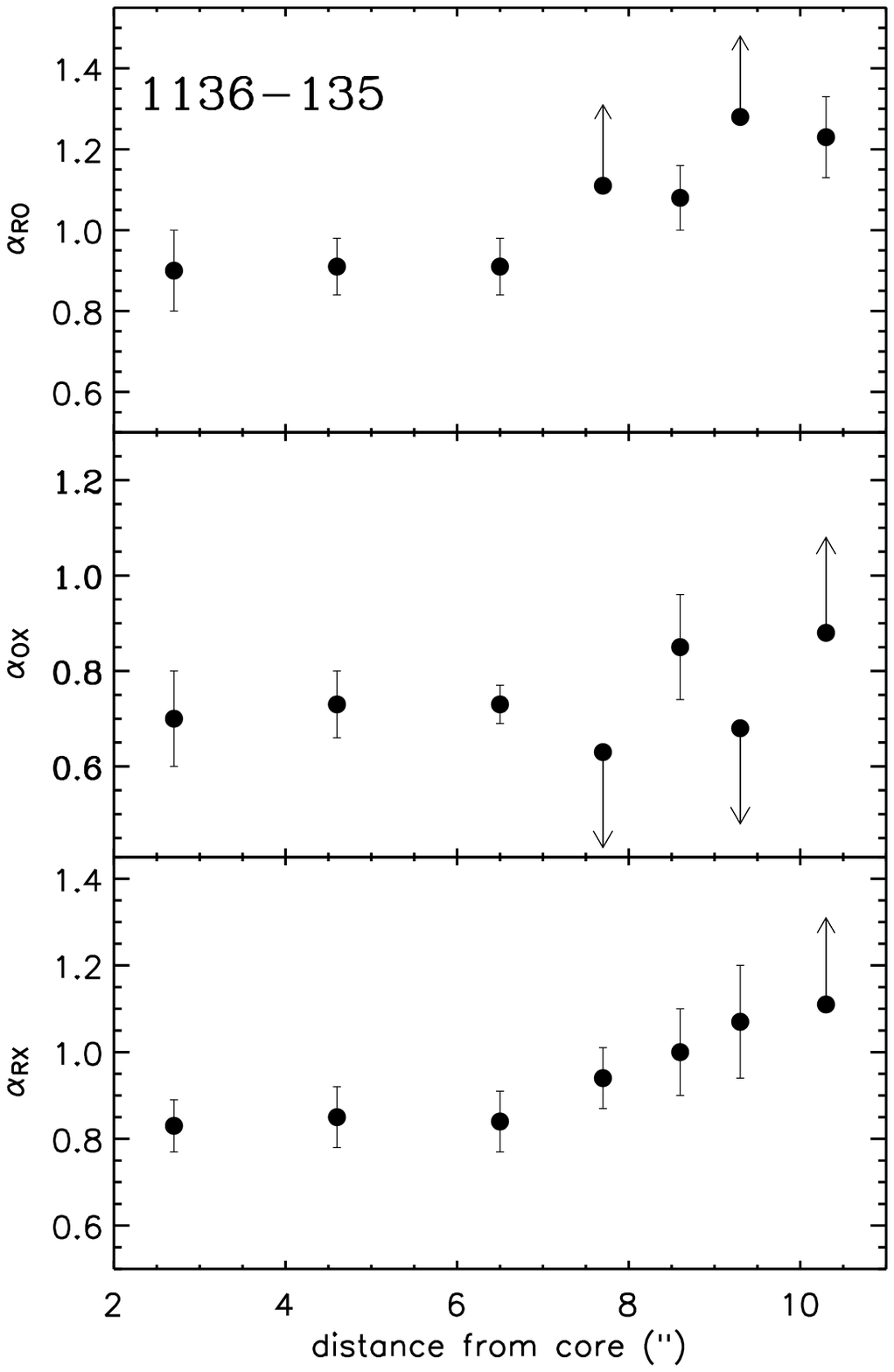}}{\includegraphics[height=9cm]{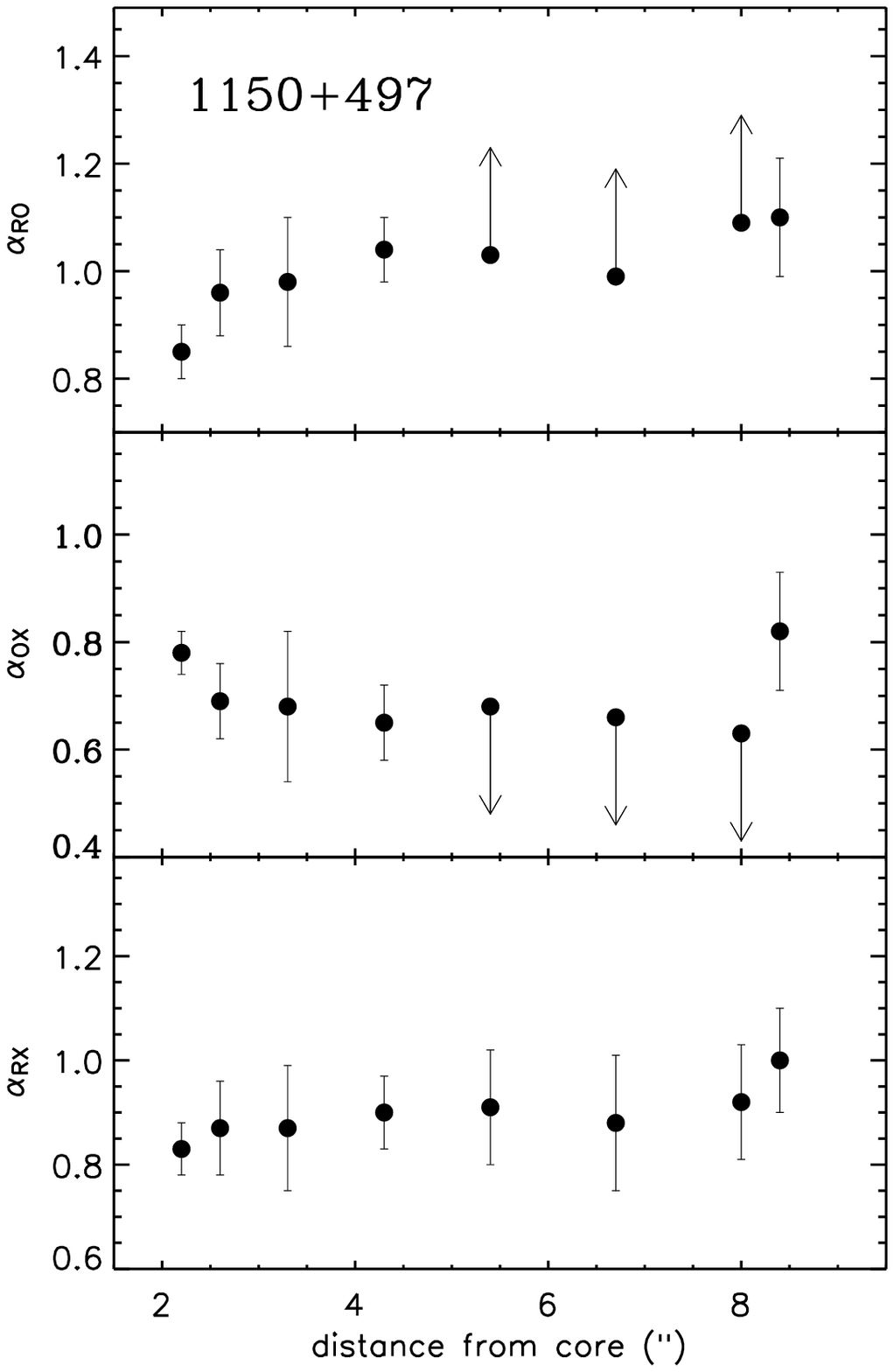}}
\vskip 0.7cm
\caption{Runs of the two-band spectral indices
$\alpha_{ro}, \alpha{ox}, \alpha_{rx}$ along the jets of 1136--135
(left) and 1150+497 (right). The radio-to-optical and
radio-to-X-ray indices increase with distance from the core in the jet
of 1136--135. For 1150+497, no demonstrable trend is present.} 
\end{figure}

\newpage

\begin{figure}
\centerline{\includegraphics[height=10cm,width=12cm]{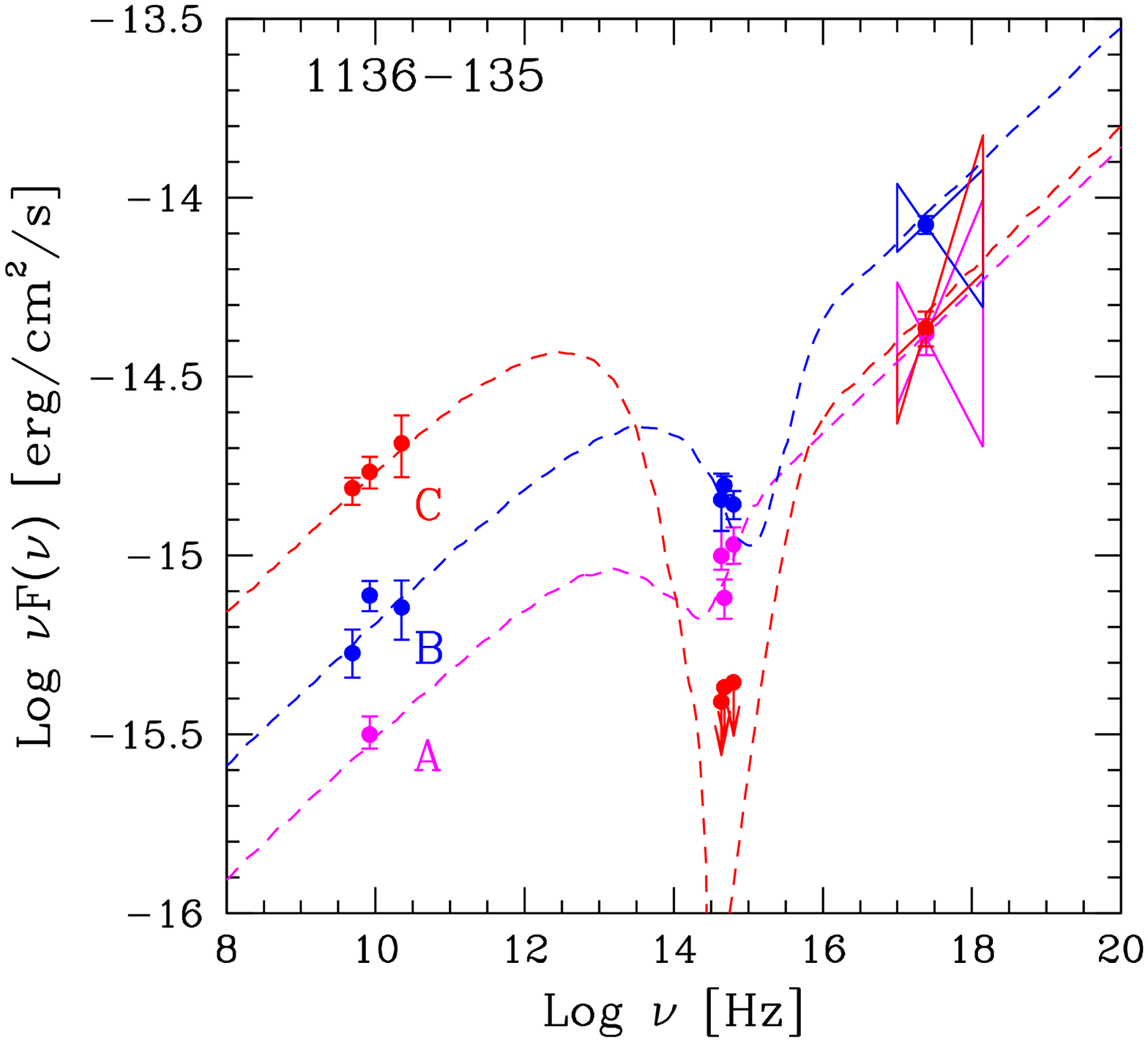}}
\vspace{0.1in}
\centerline{\includegraphics[height=10cm,width=12cm]{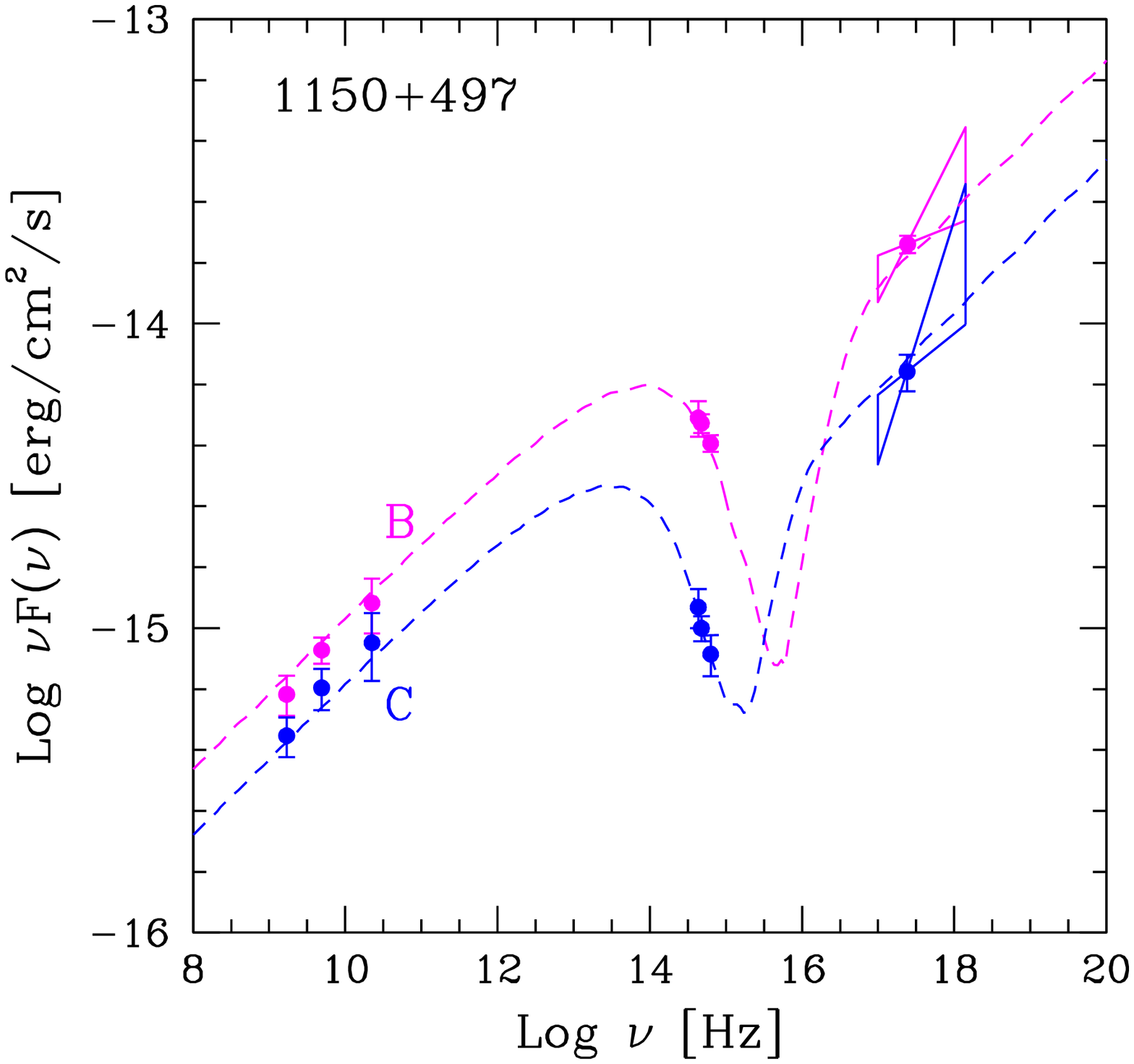}}
\caption{Spectral Energy Distributions for selected knots in the jets of 
1136--135 (top) and 1150+497 (bottom). The lines are fits with the
synchrotron+IC/CMB model (see text). Note how the optical slope is
critical to determine whether the optical emission is due to
synchrotron or is the low-energy tail of the IC/CMB component.}
\end{figure} 

\newpage 

\begin{figure}

\centerline{\includegraphics[height=10cm,width=12cm]{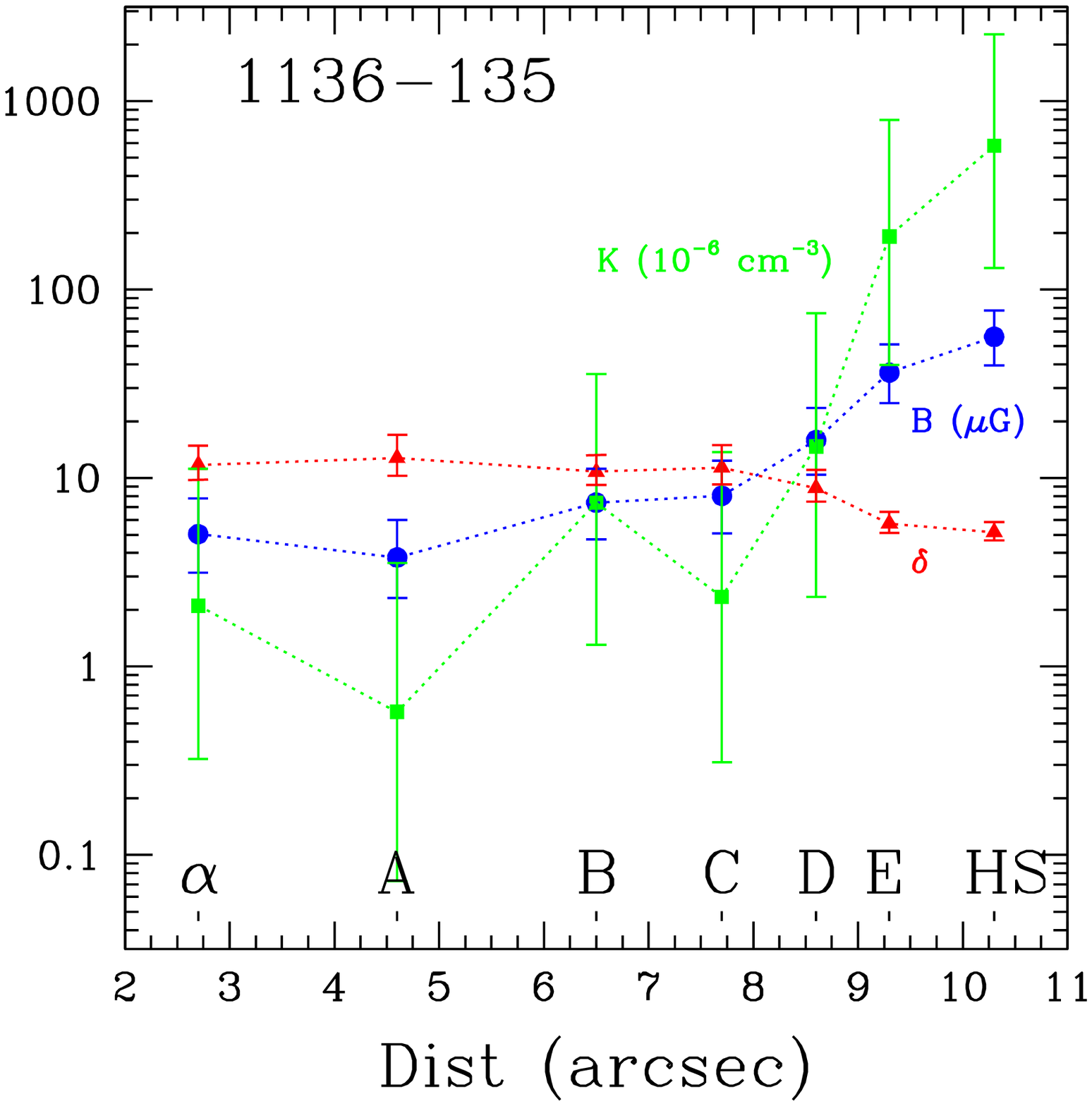}}
\vspace{0.1in}
\centerline{\includegraphics[height=10cm,width=12cm]{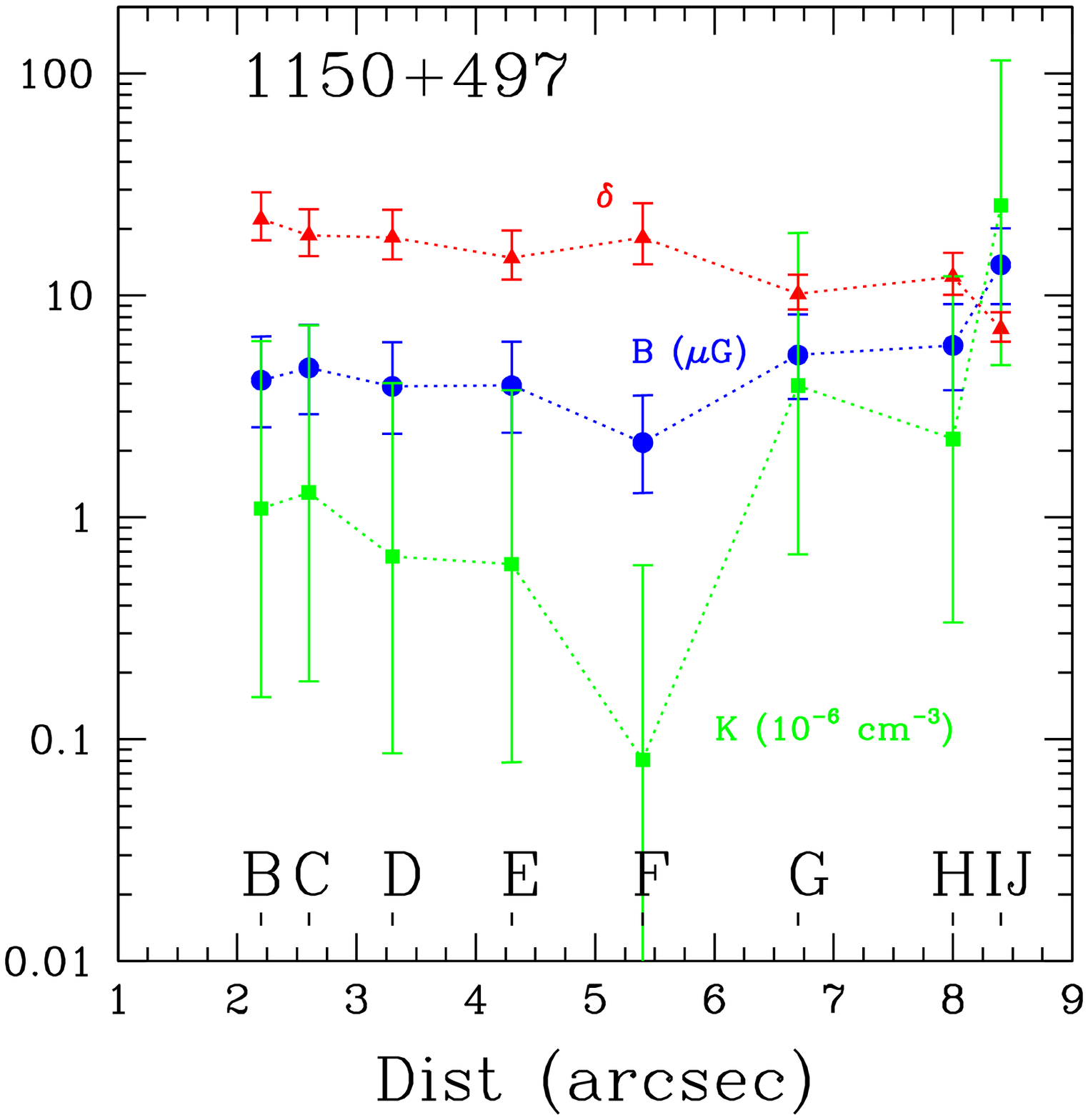}}
\caption{Plot of the Doppler factor $\delta$, magnetic field B, and
electron normalization K along the jets of 1136--135 (top) and
1150+497 (bottom). The knots positions are labeled. The parameters
were derived by fitting the IC/CMB model to the jet SEDs (see
text). In 1136--135 the parameters vary along the jet; we interpret
this as evidence for jet deceleration (see Paper II). In 1150+497,
while the Doppler factor does not change, K and B increase at the end
of the jet at the hot-spot location.}
\end{figure}

\end{document}